\documentclass{PoS}

\PoS{PoS(LAT2005)013}

\usepackage{epsfig}

\def\Vud{0.9744(4)}
\def\Vus{~~~0.2249(17)~~~}
\def\Vub{4.13(62)\!\!\times\!\! 10^{-3}}
\def\Vcd{~~0.245(22)~~}
\def\Vcs{~~0.97(10)~~}
\def\Vcb{3.91(35)\!\!\times\!\! 10^{-2}}
\def\Vtd{7.40(79)\!\!\times\!\! 10^{-3}}
\def\Vts{3.79(53)\!\!\times\!\! 10^{-2}}
\def\Vtb{0.9992(1)}

\def\Wlambda{\Vus}
\def\WA{0.77(7)}
\def\WR{0.47(8)}
\def\Wrho{0.16(7)}
\def\Weta{0.37(4)}

\def\VckmLQCDnow{V_{\rm CKM}^{\rm Lat05}}

\def\LQCDnow{{\rm Lat05}}
\def\pdg{{\rm PDG}}

\def\thisproc{these proceedings}
\def\rhob{\bar{\rho}}
\def\etab{\bar{\eta}}

\newcommand{\INT}{\mathop{\raise27pt\hbox{\bigmath Z}}}
\newcommand{\PROD}{\mathop{\raise27pt\hbox{\bigmath Y}}}
\newcommand{\SUM}{\mathop{\raise27pt\hbox{\bigmath X}}}

\let\ga=\gamma

\let\Om=\Omega

\def\to{\rightarrow}

\let\<=\langle
\let\>=\rangle

\def\beq{\begin{equation}}
\def\eeq{\end{equation}}
\def\ba{\begin{array}}
\def\bea{\begin{eqnarray}}
\def\ea{\end{array}}
\def\eea{\end{eqnarray}}
\def\bit{\begin{itemize}}
\def\eit{\end{itemize}}

\def\comment#1{ \hbox{[{\it Comment suppressed here.}\/]} }
\def\hide#1{}

\def\IR{\relax{\rm I\kern-.18em R}}
\def\IN{\relax{\rm I\kern-.18em N}}
\def\IB{\relax{\rm I\kern-.18em B}}
\def\IE{\relax{\rm I\kern-.18em E}}
\def\ZZ{\relax{\sf Z\kern-.4em Z}}     

\def\ontopss#1#2#3#4{\raise#4ex \hbox{#1}\mkern-#3mu {#2}}

\newcommand{\skipover}[1]{}
\newcommand{\nn}{\nonumber }

\def\={\!=\!}
\def\+{\,+\,}
\def\-{\,-\,}

\newcommand{\pbf}{{\bf p}}

\newcommand{\zbf}{{\bf 0}}

\newcommand{\bsl}{\begin{slide}\large}
\newcommand{\esl}{\end{slide}}

\newcommand{\GeV}{\textrm{GeV}}

\newcommand{\MeV}{\textrm{MeV}}

\newcommand{\Aerr}[2]{\ensuremath{\vphantom{0}^{+{#1}}_{-{#2}}}}

\title{Full determination of the CKM matrix using recent results
from lattice QCD}

\ShortTitle{Full determination of the CKM matrix using recent results
from lattice QCD}

\author{\speaker{Masataka Okamoto}\\
High Energy Accelerator Research Organization (KEK)\\
E-mail: \email{mokamoto@suchi.kek.jp}}

\abstract{
A full determination of the CKM matrix using recent results
from lattice QCD is presented.
To extract the CKM matrix in a uniform fashion, I exclusively
use results from unquenched lattice QCD as the theory input for 
nonperturbative QCD effects.
All 9 CKM matrix elements and all 4 Wolfenstein parameters are 
obtained from results for gold-plated quantities, which include
semileptonic decay form factors and leptonic decay constants
of $B$, $D$ and $K$ mesons,
and $B^0-\bar{B}^0$ and $K^0-\bar{K}^0$ mixing amplitudes.}

\FullConference{XXIIIrd International Symposium on Lattice Field Theory\\
                 25-30 July 2005\\
                 Trinity College, Dublin, Ireland}

\begin{document}

\section{Introduction}

The Cabibbo-Kobayashi-Maskawa (CKM) 
matrix~\cite{Cabibbo:1963yz,Kobayashi:1973fv} is 
a set of fundamental parameters in the Standard Model,
which relates the mass eigenstates and the weak eigenstates
in the electroweak theory.
In the Wolfenstein parameterization~\cite{Wolfenstein:1983yz}, 
it is given by

        \bea
        {\large V_{\rm CKM}} =
        \left(
        \begin{array}{ccc}
        { \bf{V_{ud}}} &\bf{V_{us}}  &\bf{V_{ub}}\\
        1 - \lambda^2/2     & \lambda          &A\lambda^3(\rho\!-\!i\eta)\\
          \bf{V_{cd}}  &\bf{V_{cs}}  &\bf{V_{cb}}\\
       -\lambda & 1 - \lambda^2/2 &  A\lambda^2 \\
          \bf{V_{td}}  &\bf{V_{ts}}  &\bf{V_{tb}}\\
        A\lambda^3(1\!-\!\rho\!-\!i\eta)        &  - A\lambda^2  & 1 \\
      \end{array}
        \right) .
        \eea
Because of unitarity, it contains only 4 independent parameters,
$(\lambda, A, \rho,\eta)$.

To determine each CKM matrix element, one requires both
theoretical and experimental inputs.
On the theoretical side, one needs to know relevant hadronic
amplitudes, which often contain nonperturbative QCD effects.
A major role of lattice QCD is to calculate such hadronic
amplitudes nonperturbatively, from first principles.
One can then extract the CKM matrix elements by combining
lattice QCD results as the theoretical input
with the experimental input such as branching fractions.

To accurately determine each CKM matrix element from lattice QCD,
one should use hadronic processes whose amplitude
can be reliably calculated with existent technique.
There are a set of such processes (amplitudes) ---
the so-called ``gold-plated'' processes (quantities)
which contain at most one hadron in the initial and final states~\cite{Davies:2003ik}.
These include the exclusive semileptonic decays and leptonic decays
of $B,~D$ and $K$ mesons, and neutral $B-\bar{B}$ and $K-\bar{K}$ mixings.
For such processes, technique for lattice simulations
are already well established, and thus reliable calculations
are possible.

\begin{figure}[bh]
        \bea
        \left(
        \begin{array}{ccc}
        { \bf{V_{ud}}}   &  \bf{ V_{us}}  &   \bf{ V_{ub} }\\
        \pi\to l\nu & K\to\pi l\nu  & B\to\pi l\nu \\
                    & K\to l\nu  &              \\
        \bf{ V_{cd} }  &  \bf{ V_{cs}  } &   \bf{ V_{cb}} \\
        D\to \pi l\nu & D\to K l\nu & B\!\to\! D^{(\!*\!)}\! l \nu \\
        D\to l\nu & D_s\to l\nu   \\
        \bf{ V_{td}}  &\bf{ V_{ts}}  & \bf{ V_{tb}} \\
        \langle B_d | \overline{B}_d\rangle &
        \langle B_s | \overline{B}_s\rangle \\
        \end{array}
        \right) \nn
        \eea
\caption{Gold-plated processes for each CKM matrix element~\cite{Davies:2003ik}.
The neutral $K-\bar{K}$ mixing (characterized by
the CP-violating parameter $\epsilon_K$)
is another gold-plated process, which gives a constraint on the phase of the CKM matrix
$(\rho, \eta)$.}
\label{eq:gold}
\end{figure}

The gold-plated processes for each CKM matrix element
are summarized in Fig.~\ref{eq:gold}.
The magnitude of the CKM matrix element, {\it e.g.}, $|V_{cd}|$ can be determined from either
the semileptonic decay $D\to \pi l\nu$ or leptonic decay $D\to l\nu$, as explained below.
$|V_{ub}|$ and $|V_{td}|$ can be respectively extracted from
the semileptonic $B\to \pi l\nu$ decay and neutral $B-\bar{B}$ mixing,
which give constraints on the phase of the CKM matrix $(\rho, \eta)$.
The neutral $K-\bar{K}$ mixing gives another constraint.
Taking these together, one can extract
$(\rho, \eta)$ assuming that the Standard Model is correct.
In this way one can, in principle, determine all 9 CKM matrix elements, 
{\it i.e.}, all 4 Wolfenstein parameters using lattice QCD.

The accuracy of the CKM matrix elements from lattice QCD is subject to
several systematic uncertainties, however.
The most serious one is the uncertainty from the ``quenched'' approximation,
in which effect of virtual quark loops (dynamical quarks) is neglected (``$n_f=0$'').
This approximation has been adopted in the community for a long time, simply to reduce
the computational cost.
Recent developments of computer resources and algorithms
enable us to perform more realistic lattice calculations ---
``unquenched'' simulations which include 
effect of light (up, down) dynamical quarks (``$n_f=2$'')
or light and strange dynamical quarks (``$n_f=2+1$'').
For the current status of the unquenched simulations, see Ref.~\cite{izubuchi}.

Most of gold-plated quantities listed above have been or are being calculated
in unquenched lattice QCD.
Given this situation, it would seem timely to present the CKM matrix
elements from lattice QCD in a uniform fashion in one place.
This paper gives a result for the whole CKM matrix 
-- all 9 CKM matrix elements and all 4 Wolfenstein parameters --
determined from lattice QCD using recent results for gold-plated quantities.
For this purpose I exclude results from quenched QCD. However, I note that 
quenched calculations are still important and useful to study
methodology and other systematic uncertainties.
For recent reviews on the quenched calculations of gold-plated quantities, 
see Refs.~\cite{Ryan:2001ej,Yamada:2002wh,Kronfeld:2003sd,Lubicz:2004nn,Wingate:2004xa}.

Although there are several ways to discretize quarks on the lattice,
up to now some of the gold-plated quantities have been
calculated in unquenched QCD 
only using the staggered-type fermion.\footnote{
In the staggered fermion formalism, one needs to take the fourth-root
of the fermion determinant to adjust the number of quark flavor,
as a consequence of the fermion doubling.
The validity of this procedure is not yet proven,
so further study on this issue is necessary.
For a review,
see Ref.~\cite{Durr}.}
This is because the staggered fermion
is computationally much
faster than other lattice fermion formalisms such as the Wilson-type fermion,
domain wall fermion and overlap fermion.
As a consequence, the results for the CKM matrix elements in this paper
are often estimated from only one or two unquenched calculations.
I hope that more unquenched results using other lattice fermion formalisms
will appear in the near future, 
and leave future reviewers to make a more serious average of the CKM matrix.

To present the result for the CKM matrix in a uniform fashion,
I use lattice QCD results only as the theory input for nonperturbative QCD effects.
It is of course desirable to include non-lattice theory inputs 
(such as ones for inclusive $B$ decays) to improve the precision,
but it is beyond the scope of this paper.
For recent progress on non-lattice approaches for CKM phenomenology,
see, {\it e.g.}, Refs.~\cite{ligeti,nierste,stewart}.

The results for the magnitudes of the CKM matrix elements are
        \bea \VckmLQCDnow ~=~
        \left(
        \begin{array}{ccc}
        { \bf{|V_{ud}}|}   &  \bf{ |V_{us}|}  &   \bf{ |V_{ub}| }\\
                     \Vud       &       {\Vus}    &        {\Vub} \\
        \bf{|V_{cd}| } & \bf{ |V_{cs}|} & \bf{ |V_{cb}|} \\
            {\Vcd}   &     {\Vcs}  &     {\Vcb} \\
        \bf{ |V_{td}|}  &\bf{ |V_{ts}|}  & \bf{ |V_{tb}|} \\
            \Vtd           &              \Vts     &             \Vtb  \\
      \end{array}
        \right).
\label{eq:WACKM}
        \eea
The results for the Wolfenstein parameters are
\bea
\lambda &=& \!\!\!\!\!\!\Wlambda,\\
A &=& \WA~,\\
\rhob &=& \Wrho~,\\
\etab &=& \Weta~, \label{eq:WAend}
\eea
where $\rhob = \rho (1-\lambda^2/2)$ and $\etab = \eta (1-\lambda^2/2)$.

The rest of this paper is organized as follows.
In Section~\ref{sec:magnitude}, I review recent results
for the semileptonic and letonic decays of $B,~D$ and $K$ mesons,
and determine the magnitude of the CKM matrix elements using them.
5 CKM matrix elements are directly determined using lattice results,
whereas other elements (except for $|V_{td}|$) are indirectly
determined using CKM unitarity.
In Section~\ref{sec:phase}, I review recent results
for the $B^0-\bar{B}^0$ and $K^0-\bar{K}^0$ mixing amplitudes,
and extract the CKM phase $(\rho, \eta)$ by performing 
the unitarity triangle analysis.
In Section~\ref{sec:conclusion}, I give the conclusion.

The first attempt of the full determination of the CKM matrix using lattice QCD
was made in Ref.~\cite{Okamoto:2004df}, where only results from semileptonic decays 
were used.
The determination of the CKM matrix elements using lattice QCD
has been discussed for a long time; see, {\it e.g.}, Refs.~\cite{Lubicz:2004nn,Wingate:2004xa}
for recent reviews.
A more detailed review on recent lattice calculations of $K$ meson physics
(such as the $K^0-\bar{K}^0$ mixing and $K\to\pi l\nu$ decay)
can be found in Ref.~\cite{dawson}.

\section{CKM magnitude from lattice QCD}\label{sec:magnitude}

I start this section with examples of the determination
of the magnitude of the CKM matrix elements.
The CKM magnitude is often determined from the semileptonic decays
(such as $D\to\pi l\nu$ and $K\to\pi l\nu$),
and in some cases, leptonic decays (such as $D_{}\to l\nu$ and $K\to l\nu$)
also provide an independent
determination with comparable precision. 
Below I take $|V_{cd}|$ as an example to explain how to extract
the CKM magnitude.

The branching fraction of the semileptonic decay $D\to\pi l\nu$
is given by 
\bea
\mbox{Br} (D\to\pi l\nu) ~=~  
|V_{cd}|^2  \int_{0}^{q^2_{\rm max}} dq^2 \ f_+(q^2)^2  \times (\mbox{known factor}),
\label{BrFF}
\eea
where $q = p_D - p_\pi$ is the momentum transfer,
$q_{\rm max} = (m_D-m_\pi)^2$ and 
$f_+$ is the form factor defined below.
The relevant hadronic amplitude is 
conventionally parametrized as
\bea
\langle \pi (p_\pi)|V^\mu |D(p_D)\rangle
&=&
{f_+(q^2)}
\left[p_D+p_\pi-\frac{m_D^2-m_\pi^2}{q^2}\, q\right]^\mu \nonumber
+~ f_0(q^2) \, \frac{m_D^2-m_\pi^2}{q^2} \, q^\mu ,
\label{eq:HLff}
\eea
where $V^\mu$ is the vector current involving the heavy and light quarks.
By combining the lattice calculation 
of $f_+(q^2)$ with the experimental measurement
of the branching fraction, one can extract $|V_{cd}|$.
Similarly $|V_{cs}|$, $|V_{cb}|$, $|V_{ub}|$, and
$|V_{us}|$ can be respectively extracted from 
$D\to K l\nu$, $B\to D l\nu$, $B\to \pi l\nu$ and $K\to \pi l\nu$,
as listed in Fig.~\ref{eq:gold}.

\begin{figure}[t]
\begin{center}
\leavevmode
\centerline{\hspace{-2cm}\epsfxsize=16cm \epsfbox{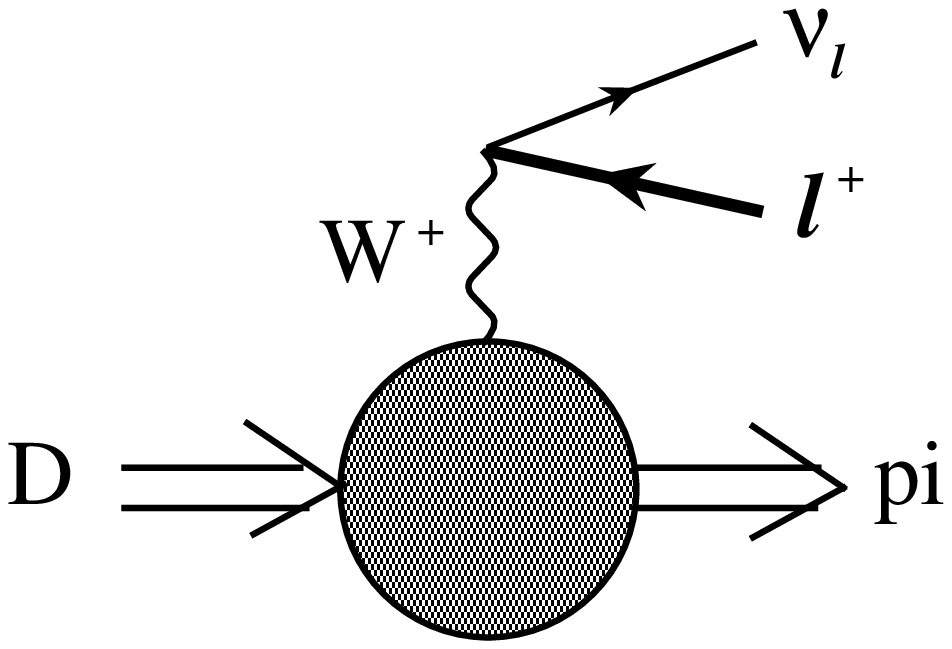}
\hspace{-7.5cm}\epsfxsize=16cm \epsfbox{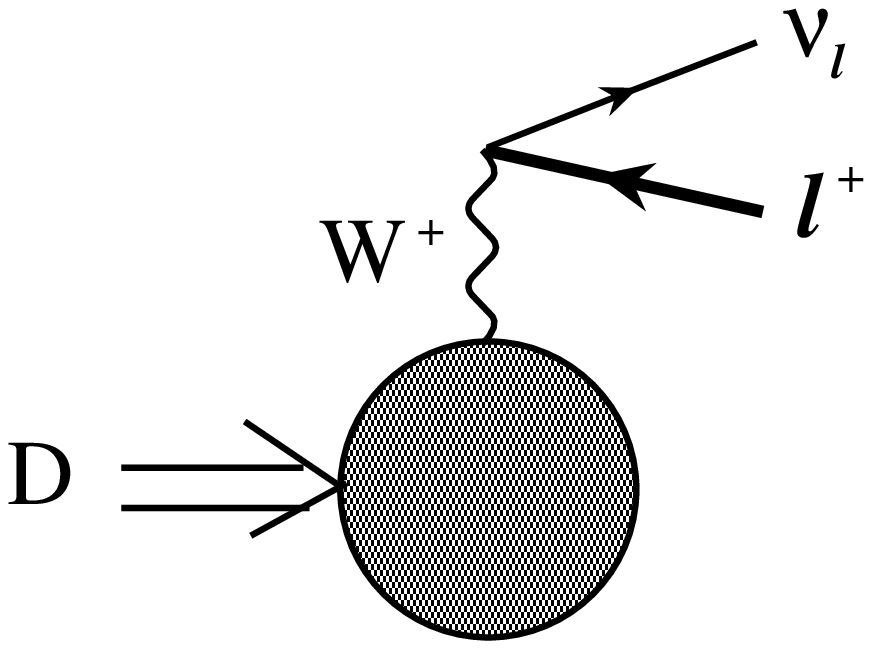}}
\vspace{-6cm}
\caption{$D$ meson decays relevant to the CKM matrix element $|V_{cd}|$.
The left panel is the semileptonic decay $D\to\pi l\nu$,
and the right is the leptonic decay $D\to l\nu$.}
\end{center}
\end{figure}

The branching fraction of the leptonic decay $D\to l\nu$ is given by
\bea
\mbox{Br} ( D\to l\nu  ) ~=~ (\mbox{known factor}) \times
 {f_D}^2 |{V_{cd}}|^2 ,
\eea
where $f_D$ is the $D$ meson decay constant defined through
\bea
\langle 0|A^\mu |D(p)\rangle
&=& {f_D} \ p^\mu.
\label{eq:fDdef}
\eea
Here $A^\mu$ is the axial vector current involving the heavy and light quarks.
Lattice results for $f_D$ can be used to determine $|V_{cd}|$
once the branching fraction is measured by experiment.
The branching fraction of $D_{}\to l\nu$ ($D_{s}\to l\nu$) decay is being (will be)
measured by CLEO-c~\cite{Bonvicini:2004gv,Artuso:2005ym,artuso} and 
that of $K \to l\nu$ has been measured by many 
groups~\cite{pdg}.
Thus they can be used to extract $|V_{cd}|$ ($|V_{cs}|$) and $|V_{us}|$
respectively.
On the other hand, $B\to l\nu$ decay is still difficult to measure 
since the branching fraction is suppressed by a small factor $|V_{ub}|^2$.

\subsection{$|V_{cd}|$ and $|V_{cs}|$ from $D$ meson decays}

There are (at least) two reasons to study the $D$ meson decays in lattice QCD.
One is to extract the CKM matrix elements, as mentioned above.
The other is that,
taken the CKM matrix elements from elsewhere,
one can test lattice QCD by comparing lattice results
for the form factor $f_+^{D\to\pi(K)}$ and decay constant 
$f_D$ with corresponding experimental results.
In particular,
if lattice results agree with experiment for the $D$ meson physics, one can have confidence
in similar quantities for the $B$ meson physics such as
$f_+^{B\to\pi}$ and $f_B$, which are phenomenologically important and cannot
be obtained by other means.

\begin{figure}[h]
\begin{center}
\leavevmode
    \epsfxsize=10cm
\centerline{\epsfxsize=12cm \epsfbox{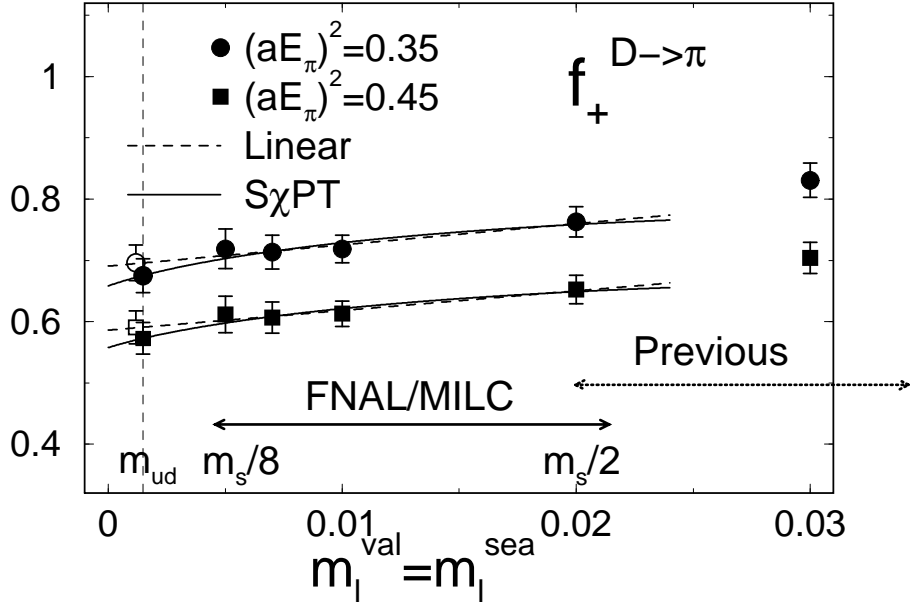}}
\caption{Light quark mass dependence of the $D\to\pi$ form factor in Ref.~\cite{Aubin:2004ej}.
Symbols are data points, and lines are chiral fits using the staggered chiral perturbation
theory (solid lines) and ones using a linear ansatz (dashed lines). The dashed vertical line
indicates the physical $ud$ quark mass.
}
\label{mdepff}
\end{center}
\end{figure}

\begin{figure}
\epsfig{file=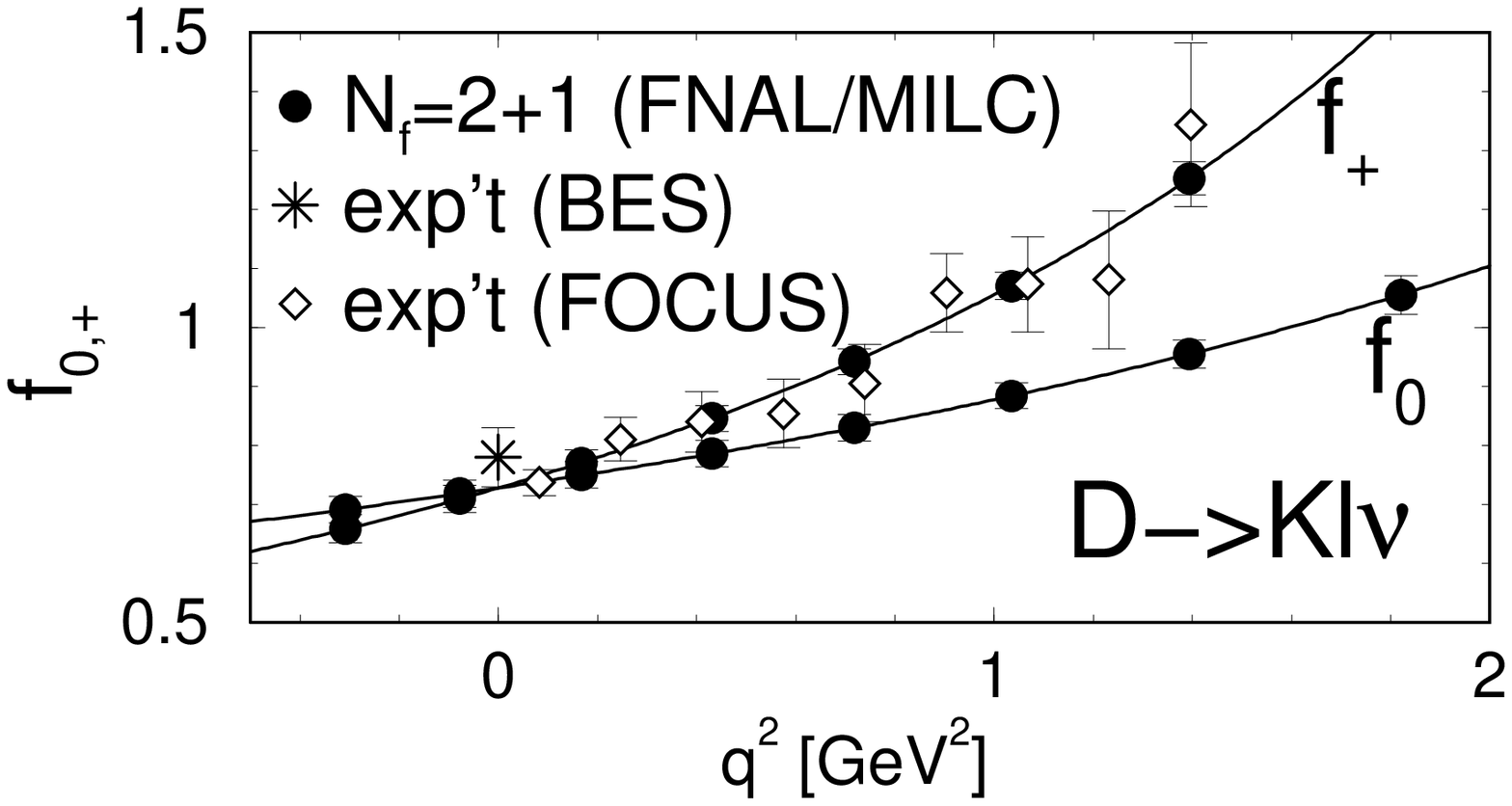, width=.8\textwidth, height=.3\textheight}
\epsfig{file=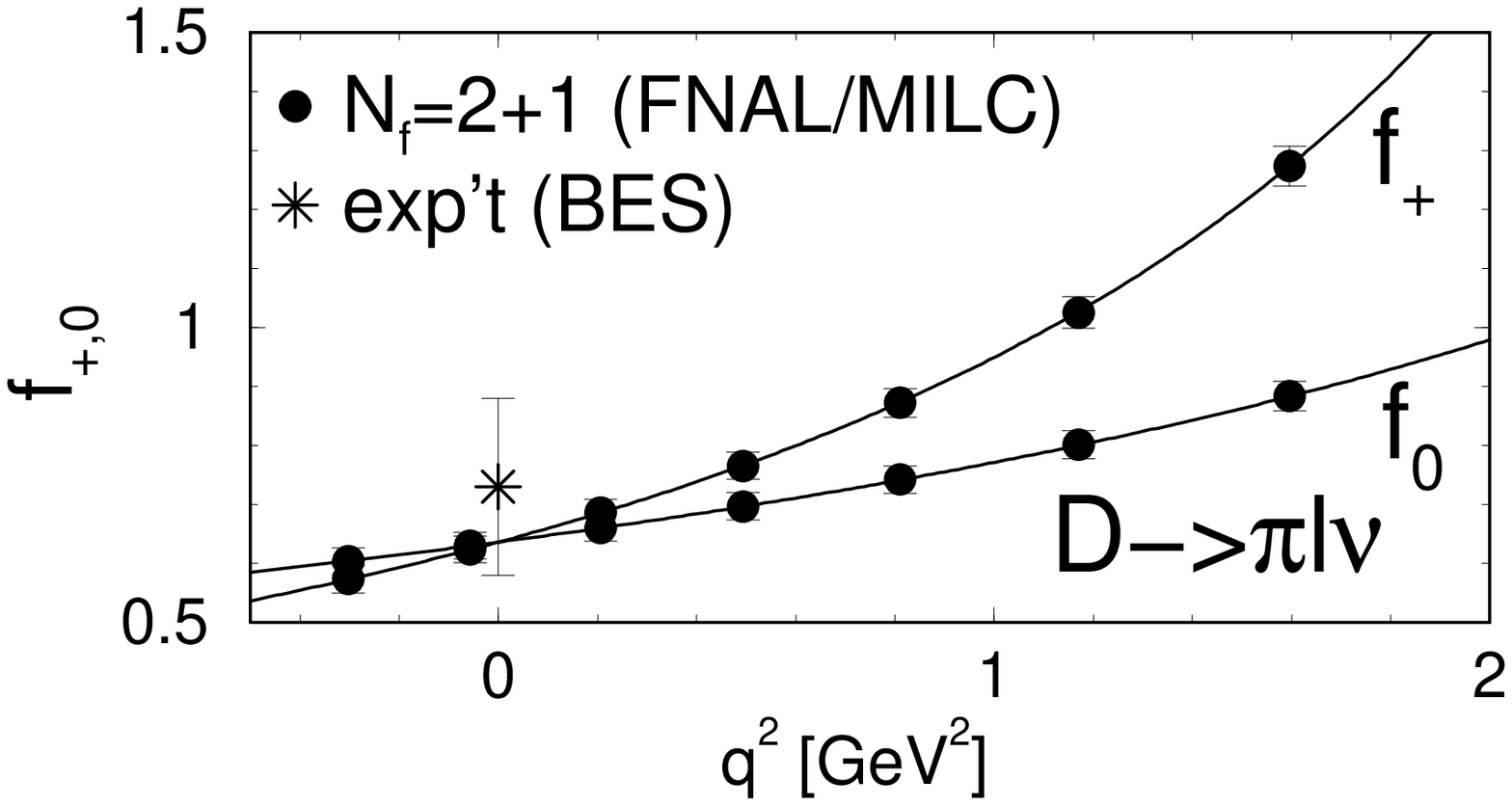, width=.8\textwidth, height=.3\textheight}
\caption{Form factors of $D\to\pi l \nu$ and $D\to K l \nu$
decays in $n_f=2+1$ lattice QCD by the Fermilab/MILC collaboration~\cite{Aubin:2004ej}.
The errors are statistical only.
Recent experimental results by the BES collaboration~\cite{unknown:2004nn} and by the FOCUS
collaboration~\cite{Link:2004dh} are also shown.} 
\label{fig1}
\end{figure}

The Fermilab/MILC collaboration presented the first unquenched ($n_f=2+1$)
calculation of form factors of $D\to\pi l \nu$ and $D\to K l \nu$
decays~\cite{Aubin:2004ej,Okamoto:2004xg}
using the MILC ``coarse lattice'' ensemble ($a^{-1}\approx 1.6~\GeV$)~\cite{milc}.
We use an improved staggered fermion action~\cite{asqtad} for light ($u,d,s$) quarks 
and an improved Wilson fermion action~\cite{Sheikholeslami:1985ij} 
with the Fermilab interpretation~\cite{El-Khadra:1996mp} for the charm quark. 
To combine the staggered-type light quark (1-component spinor) with the Wilson-type
heavy quark (4-component spinor), 
we convert the staggered quark propagator into the naive quark 
(4-component spinor) propagator according to
\bea
\Om(x)^\dagger \<\bar{\chi}(x)\chi(y)\>\Om(y) = \<\bar{\psi}(x)\psi(y)\> ,
\eea
where $\Om(x) = \ga_0^{x_0}  \ga_1^{x_1}  \ga_2^{x_2}  \ga_3^{x_3}$,
as proposed in Refs.~\cite{Wingate:2002fh,lepage}.
Although the naive quark propagator describes
16 (=$2^4$) equivalent fermions, known as the ``doubling problem'',
only the physical mode contributes to the low energy physics 
and the remaining 15 ($=16-1$) modes decouple
when it is combined with the Wilson-type (or any doubler-free) heavy quark.
This can be understood by noting that the heavy-light meson mass
for each mode is roughly given by 
$M_D^{} \simeq \{m_u+m_c,~ m_u+(m_c+2r/a),~ m_u+(m_c+4r/a), \cdots\}$
with $r$ being the Wilson parameter for the heavy quark action.
By taking the asymptotic limit 
in the time direction ($t\to \infty$) for heavy-light 2-point and 3-point functions,
the state with the lowest mass (physical mode) can be isolated.
This method
has been successfully applied to the heavy-light meson physics
by the HPQCD collaboration and Fermilab/MILC collaboration.\footnote{
Note that this method may be applied to the $K$ meson physics as well,
with the staggered $u,d$ quarks and the Wilson-type $s$ quark.}

Since the staggered fermion is fast and free from the exceptional configurations,
one can perform simulations with the light quark mass $m_l$ as low as $m_s/8$,
in contrast to previous calculations using other fermions.
This situation is shown in Fig.~\ref{mdepff}.
Physical results at $m_l=m_{ud}\equiv (m_u+m_d)/2$ 
are obtained from chiral fits using 
the staggered chiral perturbation theory (S$\chi$PT)~\cite{schiPT1,schiPT2}, but
a simple linear fit gives consistent results.
This suggests that the results at the physical light quark mass
are insensitive to the fit ansatz
and the error from the chiral extrapolation ($m_l\to m_{ud}$)
is under control. The availability of data at small light quark masses is crucial.
On the other hand, since our calculation is done at a single lattice spacing,
the error from the lattice discretization effects is still large,
giving about 10\% total systematic uncertainties for the form factors.

Our final results are shown in Fig.~\ref{fig1}
together with recent experimental results~\cite{unknown:2004nn,Huang:2004fr,Link:2004dh,:2005sh}.
The results are obtained 
using the parameterization of Becirevic and Kaidalov (BK)~\cite{Becirevic:1999kt},
\bea\label{eq:BK}
f_+(q^2) = \frac{F}{(1-\tilde{q}^2)(1-\alpha\tilde{q}^2)},~~~
f_0(q^2) = \frac{F}{1-\tilde{q}^2/\beta},
\eea
where $\tilde{q}^2=q^2/m_{D^{*}}^2$.
We obtain~\cite{Aubin:2004ej,Okamoto:2004xg}
\bea
F^{D\to \pi}=0.64(3)(6),~&
\alpha^{D\to  \pi}=0.44(4)(7),~& \beta^{D\to \pi}=1.41(6)(13), \\
F^{D\to  K}=0.73(3)(7),~  &
\alpha^{D\to  K}=0.50(4)(7),~  & \beta^{D\to  K}=1.31(7)(13),
\eea
where the first errors are statistical and the second
systematic.
The results agree with experiment for both
the normalization at $q^2=0$~\cite{unknown:2004nn,Huang:2004fr} and 
the $q^2$-dependence~\cite{Link:2004dh,:2005sh}.
This may indicate reliability of lattice results for the similar quantities
for $B$ physics, the $B\to\pi l\nu$ form factor.
By integrating out $f_+(q^2)$ in terms of $q^2$ and using 
the experimental measured branching fraction~\cite{pdg},
we obtain 
\bea
   |V_{cd}|_{\rm semi-lep} & = & 0.239(10)(24)(20),  \label{eq:Vcd} \\
   |V_{cs}|_{\rm semi-lep} & = & 0.969(39)(94)(24), \label{eq:Vcs}
\eea
where the first errors are statistical,
the second systematic, and the third are the experimental errors from
the branching fractions.

\begin{figure}[tb]
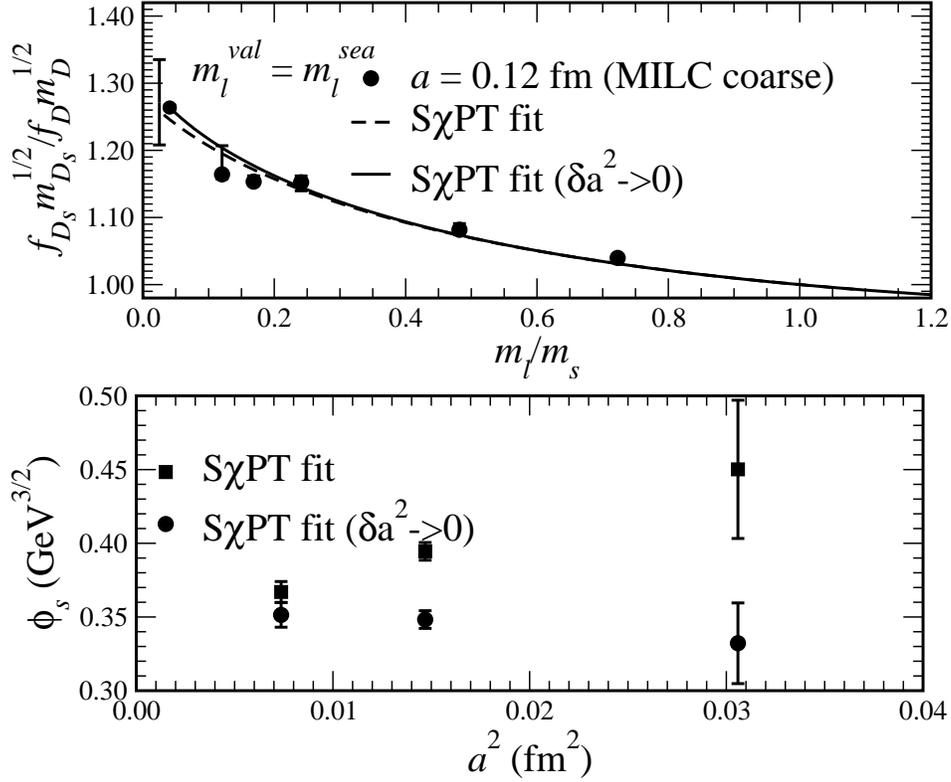

\begin{center}
\leavevmode
    \epsfxsize=10cm
\centerline{\epsfxsize=12.5cm \epsfbox{xi-mq.proc.eps}}
\centerline{\epsfxsize=12.5cm \epsfbox{phiDs-a2-fm.proc.eps}}
\caption{$D_{}$ and $D_{s}$ meson decay constants in $n_f=2+1$ lattice QCD 
by the Fermilab/MILC collaboration~\cite{Aubin:2005ar}.
The upper figure shows the light quark mass dependence
and chiral extrapolation for $f_{D_s}\sqrt{m_{D_s}}/(f_{D}\sqrt{m_{D}})$.
The lower figure shows the lattice spacing dependence of 
$\phi_s = f_{D_s}\sqrt{m_{D_s}}$.}
\label{fnalfD}
\end{center}
\end{figure}

The Fermilab/MILC collaboration has also
finalized the $n_f=2+1$ calculation
of the $D_{}$ and $D_{s}$ meson decay constants~\cite{Aubin:2005ar,simone,kronfeld}.
The employed lattice actions are the same as the ones for 
the semileptonic calculations. 
We performed partially quenched simulations,
where the valence light quark mass can be different from
the dynamical light quark mass,
at three values of the lattice spacing.
The chiral extrapolation is done using the S$\chi$PT formula~\cite{Aubin:2005aq}, and 
the final results are obtained by taking
$\delta a^2\to 0$ after the chiral extrapolation (upper figure of Fig.~\ref{fnalfD}),
where $\delta a^2$ denotes constants in the S$\chi$PT which parametrizes
lattice discretization effects from the staggered fermion.
After $\delta a^2\to 0$, the lattice spacing dependence is small (lower figure
of Fig.~\ref{fnalfD}). Our final results are
$f_{D} = 201(03)(17)\;\MeV$ and
$f_{D_s} = 249(03)(16)\;\MeV$,
where the first errors are statistical and the second systematic.
The two largest sources of the systematic error are 
the discretization effects and the chiral extrapolation (for $f_{D}$).

\begin{figure}[tb]
\begin{center}
\leavevmode
    \epsfxsize=10cm
\centerline{\epsfxsize=13cm \epsfbox{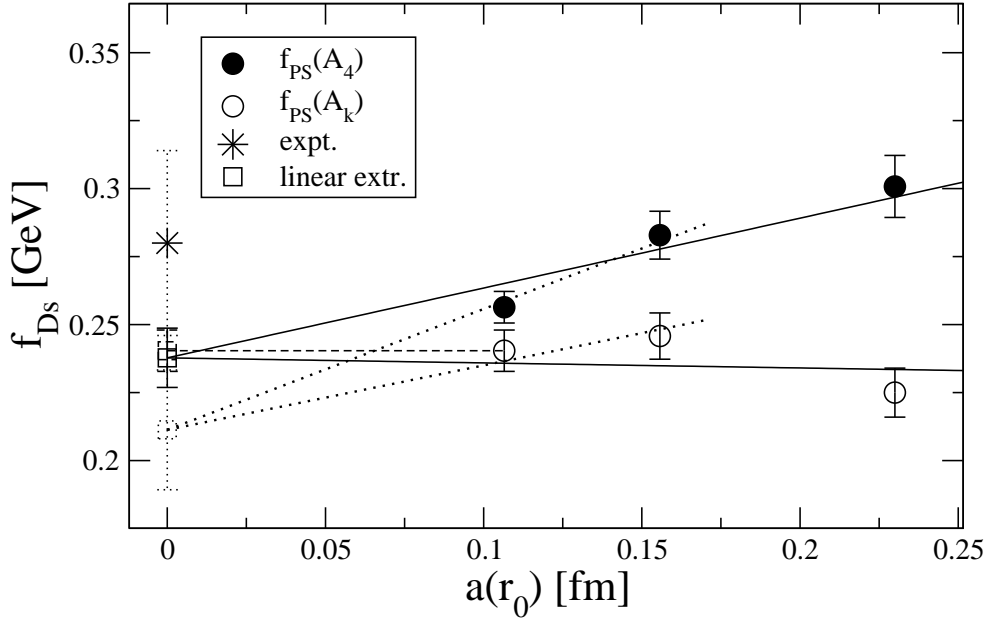}}
\caption{Lattice spacing dependence and continuum extrapolation ($a\to 0$)
for $f_{D_s}$ in $n_f=2$ lattice QCD 
by the CP-PACS collaboration~\cite{kuramashi,cppacs_fd}.
Filled circles are results from the temporal axial vector current and 
open circles are from the spatial one.
}
\label{fig:cppacs_fd}
\end{center}
\end{figure}

The CP-PACS collaboration reported a new unquenched ($n_f=2$) calculation 
of the $D_{(s)}$ decay constant~\cite{kuramashi,cppacs_fd}.
For the light quarks, they used an $O(a)$ improved Wilson action.
For the charm quark, they used a relativistic on-shell improved 
action~\cite{Aoki:2001ra,Aoki:2003dg}
which is similar to one in Ref.~\cite{El-Khadra:1996mp} 
but derived from a different point of view in Ref.~\cite{Aoki:2001ra}.
Their simulated light quark mass ranges $m_l\ge m_s/2$,
and a linear chiral extrapolation was made.
They studied the lattice spacing dependence using three lattice spacings
as in the Fermilab/MILC calculation, and performed a continuum extrapolation
combining two data sets (one is from the temporal axial vector current and the other
is from the spatial current)
as shown in Fig.~\ref{fig:cppacs_fd}.
Their preliminary results are
$f_{D} = 202(12)(\Aerr{20}{25})\;\MeV$ and
$f_{D_s} = 238(11)(\Aerr{07}{27})\;\MeV$,
where the systematic errors (second parentheses) are dominated by
uncertainties from the continuum extrapolation.

The chirally improved actions were applied to the charm quark physics
in two recent quenched ($n_f=0$) studies.
Ref.~\cite{Chiu:2005ue} used the overlap fermion action, whereas
Ref.~\cite{dong} used a variant of the domain-wall fermion action.
The use of the chirally improved actions has an advantage that 
they are free from $O(a,m_Qa)$ discretization errors without any tuning of parameters, 
and so may be useful for the simulations involving the charm quark where $m_Qa < 1$.

Turning to the experimental result for the leptonic decay,
the CLEO-c collaboration updated their measurement of $D\to\mu\nu$
branching fraction~\cite{Artuso:2005ym,artuso}.
Assuming that $|V_{cd}|=|V_{us}|=0.224(3)$, they obtain the experimental result
$f_{D}^{\rm CLEO-c} = 223(17)(03)\;\MeV$,
which is in agreement with the recent lattice results.
Their precision, $O(10\%)$, is similar to that for the lattice results.
The agreement may indicate reliability of lattice calculations of the
heavy-light decay constants.
They will further improve the precision of  $f_{D}$ and report
the result for $f_{D_s}$ in the future.

One may conversely use the CLEO-c result for the $D\to\mu\nu$
branching fraction to determine $|V_{cd}|$.
If it is combined with the $n_f=2+1$ lattice result for $f_{D}$ by the Fermilab/MILC
collaboration, one obtains
\bea
|V_{cd}|_{\rm lept}=0.250(22)(21),
\label{eq:Vcd_lept}
\eea
where the first error is one from the lattice calculation and the second is 
from the experimental uncertainty. 
$|V_{cd}|$ from the leptonic decay is consistent with the value obtained
from the semileptonic decay
Eq.~(\ref{eq:Vcd}), and the size of uncertainties is similar to each other.

It is interesting to consider the ratio of 
the leptonic and semileptonic decay branching fractions
because the CKM matrix element $|V_{cd}|$ cancels in the ratio.
Writing
\bea
R \equiv \sqrt{\frac{{\rm{Br}}( D\to  l\nu  ) }{{\rm{Br}} ( D \to \pi l\nu  )}},
\eea
the $R$ is proportional to ${f_{D}}/[{\int dq^2 (f^{D\to\pi}_+(q^2))^2}]^{1/2}$.
Since one can directly compare lattice results with experiment for this quantity,
it provides a good test of lattice QCD.
The experimental result (dominated by the CLEO-c measurements) is~\cite{artuso}
\bea
R_{\rm exp} = 0.25(2),
\eea
and the unquenched lattice result using 
the Fermilab/MILC calculation of $f_{D}$ and  $f^{D\to\pi}_+$
is 
\bea
R_{\rm lat} = 0.21(3).
\eea
They 
are in reasonable agreement with 
each other.

As for the Lattice'05 value of the CKM matrix elements,
I take a weighted average of Eqs.~(\ref{eq:Vcd}) and (\ref{eq:Vcd_lept})
for $|V_{cd}|$, and simply quote Eq.~(\ref{eq:Vcs}) for $|V_{cs}|$;
\bea
   |V_{cd}|_{\LQCDnow} & = & \Vcd,  \label{eq:VcdWA} \\
   |V_{cs}|_{\LQCDnow} & = & \Vcs, \label{eq:VcsWA}
\eea
where the errors are the combined uncertainties from
theory (from lattice QCD) and experiment.
These are consistent with the values quoted in Particle Data
Group (PDG)~\cite{pdg}; 
$|V_{cd}|_{\pdg}=0.224(12)$ and $|V_{cs}|_{\pdg}=0.996(13)$.
Note that the PDG values above are obtained
from neither the semileptonic nor the leptonic decays, and thus
Eqs.~(\ref{eq:VcdWA}) and (\ref{eq:VcsWA}) provide an independent
determination of $|V_{cd}|$ and $|V_{cs}|$.

\subsection{$B$ meson decays}

\subsubsection{$B\to \pi l\nu$ decay}

The semileptonic decay $B\to\pi l\nu$ can be used to determine
$|V_{ub}|$, which is one of the most important CKM parameters to constrain the unitarity triangle.
Since the pion momentum available in lattice calculations is limited up to
around $1 ~\GeV$, only the higher $q^2$-region ($q^2\ge 16 ~\GeV^2$) can be simulated
for the $B\to\pi l\nu$ decay. This is in contrast to 
the $D\to\pi l\nu$ decay, for which we can cover all $q^2$-region ($0 \le q^2\le q^2_{\rm max}$).
Since the experimental accuracy of the branching fraction is better
for the lower $q^2$-region than the higher $q^2$-region, it is desirable to
extend the lattice result to the lower $q^2$-region.
Below I first summarize recent lattice results for the $B\to\pi l\nu$ form factors,
and extract $|V_{ub}|$ using the results for the higher $q^2$-region.
I then discuss recent attempts to extend the results to the lower $q^2$-region.

\begin{figure}[t]
\begin{center}
\leavevmode
\includegraphics*[width=13cm]{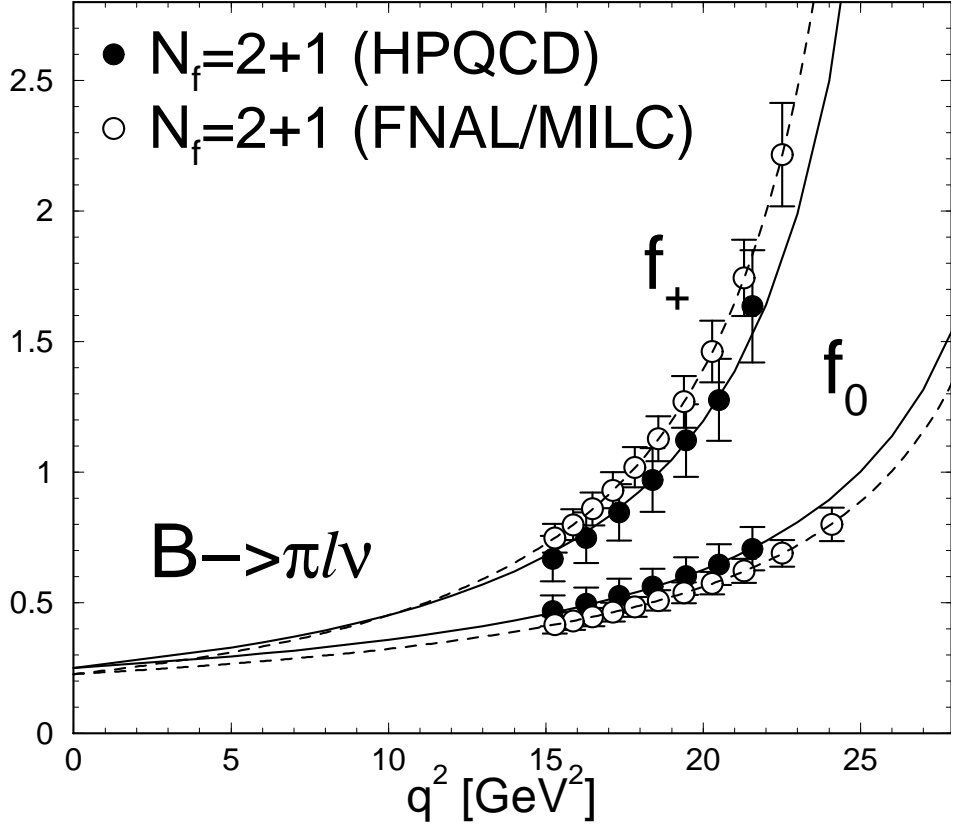}
\caption{
Form factors of the $B\to\pi l \nu$ decay in $n_f=2+1$ lattice QCD 
by the HPQCD collaboration~\cite{Shigemitsu:2004ft,gulez} and 
by the Fermilab/MILC collaboration~\cite{Okamoto:2004xg,mackenzie}.
Lines are ones from fits with the BK parameterization.
}
\label{fig:b2pi}
\end{center}
\end{figure}

There are two $n_f=2+1$ unquenched calculations of the $B\to\pi l\nu$ form factors
using the MILC coarse lattice
ensembles with improved staggered light quarks.
One is obtained with the NRQCD heavy quark~\cite{Lepage:1992tx}
by the HPQCD collaboration~\cite{Shigemitsu:2004ft,gulez}, and
the other is with the Fermilab heavy quark
by the Fermilab/MILC collaboration~\cite{Okamoto:2004xg,mackenzie}.
Their results are shown in Fig.~\ref{fig:b2pi}.
The total systematic uncertainties are 11\% for the form factors in both calculations.
The largest quoted 
error comes from the perturbative matching between the lattice and continuum theory
for the HPQCD result, and from the discretization effect for the Fermilab/MILC result.
Two unquenched results agree with each other within errors.
These are also consistent with previous quenched 
calculations~\cite{El-Khadra:2001rv,Abada:2000ty,Aoki:2001rd,Bowler:1999xn,Shigemitsu:2002wh};
at present it is difficult to estimate the effect of the quenching quantitatively
because the size of other systematic uncertainties, $O(10\%)$, is the same as 
the expected size of the quenching error.

By integrating $f_+(q^2)$ over $16\ \GeV^2 \le q^2\le q^2_{\rm max}$
and using an average~\cite{HFAG} of 
the partial branching fractions $\mbox{Br}(16\ \GeV^2 \le q^2\le q^2_{\rm max})$
measured by CLEO~\cite{Athar:2003yg}, Belle~\cite{Belle:B2pi,Abe:2005ie,Belle:eps05}
 and BABAR~\cite{Aubert:2005tm,Aubert:2005cd,Aubert:2005sb} collaborations, 
the CKM matrix element is obtained as  
\bea
|V_{ub}|_{\rm HPQCD}~~~~~~ & = & 4.47~(22)(49)(30)     \!\times\! 10^{-3}\label{Vubhpqcd},\\
|V_{ub}|_{\rm FNAL/MILC} & = & 3.78~(30)(42)(25) \!\times\! 10^{-3}\label{Vubfnal},
\eea
where the first errors are statistical,
the second systematic, and the third are the experimental errors.
As for the Lattice 2005 value, I take a simple average of the two preliminary results,
obtaining
\bea
|V_{ub}|_{\LQCDnow} & = & \Vub.  \label{eq:VubWA} 
\eea
The total uncertainty for $|V_{ub}|$ is 15\%.

\begin{figure}[t]
\begin{center}
\leavevmode
    \epsfxsize=10cm
\centerline{\hspace{0cm}\epsfxsize=12cm \epsfbox{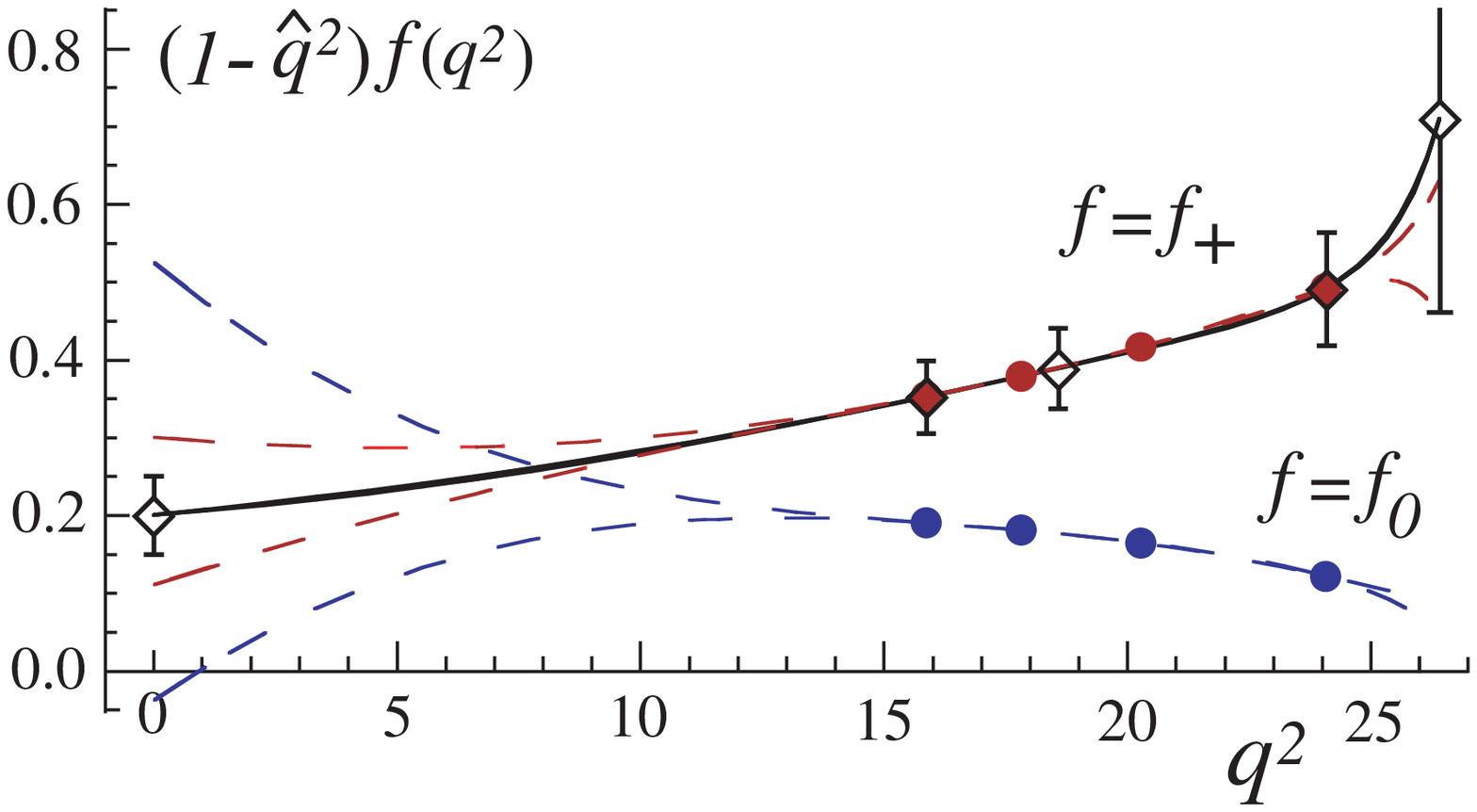}}
\caption{
$B\to\pi l\nu$ form factors using the unquenched lattice results,
QCD dispersion relation and $|V_{ub}|f_+(0)$ from an analysis
based on the soft collinear effective 
theory~\cite{Arnesen:2005ez}.
}
\end{center}
\end{figure}

Let us discuss the $q^2$-dependence of the $B\to\pi l\nu$ form factors.
Both HPQCD and Fermilab/MILC collaborations use the BK
parameterization Eq.~(\ref{eq:BK}) to 
interpolate and extrapolate the results in $q^2$.
To estimate the
uncertainty from the BK parameterization,
the Fermilab/MILC collaboration also made a fit using a
polynomial ansatz in $q^2$.
The difference between the two methods for
$|V_{ub}| \propto (\int_{q^2_{\rm min}}^{q^2_{\rm max}} dq^2 f_+(q^2)^2)^{-1/2}$
is 4\% with $q^2_{\rm min}=16~\GeV^2$.
This difference is included in the systematic error in Eq.~(\ref{Vubfnal}).
With $q^2_{\rm min}=0$, however, the difference amounts to be 11\%,
which is a significant effect.
This is because a long extrapolation is required
from lattice data points ($16~\GeV^2 \le q^2 \le q^2_{\rm max}$) to $q^2=0$.
For a more precise determination of $|V_{ub}|$, 
it is necessary to reduce the uncertainty for the lower $q^2$-region.

One solution is to combine the lattice results for the higher $q^2$-region
with non-lattice results for the lower $q^2$-region. 
Ref.~\cite{Arnesen:2005ez} combined the recent unquenched lattice results with
the QCD dispersion relation and $|V_{ub}|f_+(0)$ from an analysis of the $B\to\pi\pi$ decay 
 based on the Soft Collinear Effective Theory
(SCET). 
The QCD dispersion relation is not a model but an analyticity bound.
Ref.~\cite{Albertus:2005ud} 
made a similar study using $f_+(0)$ from the light-cone sum rule
instead of one from the SCET. 
Ref.~\cite{Fukunaga:2004zz} combined quenched lattice results with
the QCD dispersion relation and the experimental measured $q^2$-dependence.
In each case, the uncertainty for 
$|V_{ub}|$ can be reduced by $\approx$5\% with the additional information on the form factors.

\begin{figure}[t]
\begin{center}
\leavevmode
\includegraphics*[width=10.5cm]{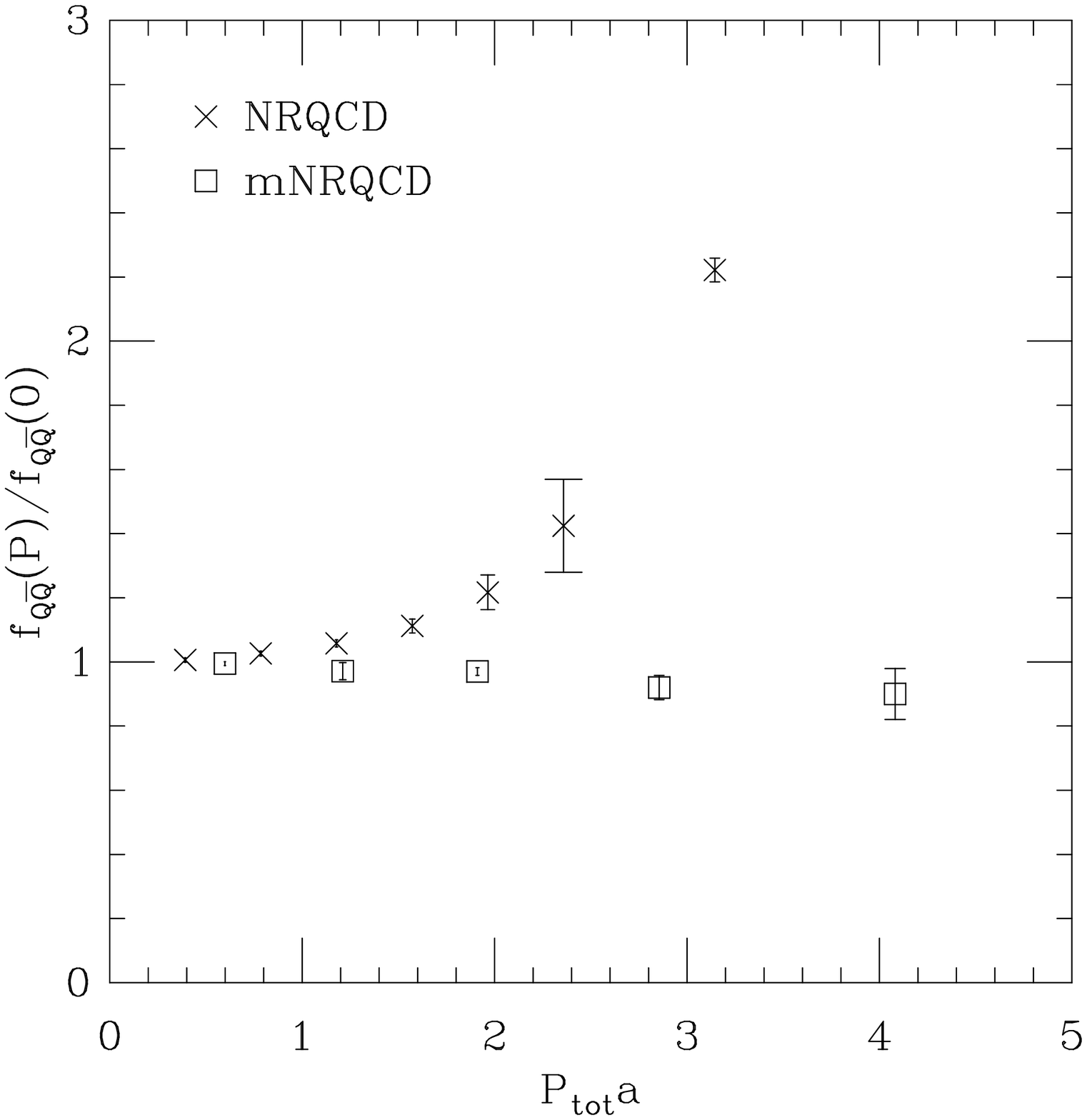}
\caption{Momentum-dependence of the ratio of 
heavy-heavy meson decay constants $f_{HH}(\pbf)/f_{HH}(\zbf)$
from NRQCD (crosses) and mNRQCD (squares)~\cite{dougal,davies}.}
\label{fig:mNRQCD}
\end{center}
\end{figure}

Another solution is a direct lattice simulation 
at lower $q^2$ using the moving NRQCD (mNRQCD)~\cite{Hashimoto:1996xq,Sloan:1997fc,Foley:2002qv}.
The mNRQCD is a generalized version of non-relativistic
QCD in the $B$ meson moving frame 
(${\bf u} = \pbf_B/M_B {~\ne~} {\bf 0}$). The action is given by
\bea
{\cal L}_{\rm mNRQCD} = \psi^\dagger
\left(i D_t + { i ({\bf v\cdot D})} + \frac{{\bf D}^2}{2\gamma m}
 - { \frac{({\bf v\cdot D})^2}{2\gamma m}} + \cdots \right)
\psi
\eea
where $u_\mu = \gamma (1,{\bf v})$ and $\gamma^{-1} = \sqrt{1-{\bf v}^2}$.
Setting ${\bf v }={\bf 0}$ gives the usual NRQCD action.
The mNRQCD allows the $B\to\pi l\nu$ calculation at lower $q^2$ with
smaller $\pbf_\pi$;
{for ${\bf v}\approx 0.75$,} $q^2=0$ can be achieved with $\pbf_\pi=1~\GeV$~\cite{hashimoto},
where the size of lattice discretization effects from non-zero pion momentum
is modest.

Previous lattice simulations with the mNRQCD for a large ${\bf v}$
suffered from large statistical errors~\cite{Hashimoto:1996xq},
but Ref.~\cite{Foley:2004rp} showed that it can be reduced by using a special smearing function 
so that the simulation with ${\bf v}\approx 0.7$--0.8 is feasible.
Ref.~\cite{dougal} made a quenched study to test the mNRQCD.
They calculated the decay constant of a heavy-heavy meson $f_{HH}$ at non-zero
momentum $\pbf$ using both the NRQCD (${\bf v }={\bf 0}$) action
and mNRQCD (${\bf v }\ne{\bf 0}$) action.
As shown in Fig.~\ref{fig:mNRQCD},
$f_{HH}$ with the NRQCD depends on $\pbf$, indicating that 
the discretization effect from non-zero pion momentum is not under control.
On the other hand, 
the result with the mNRQCD is constant in $\pbf$ as it should be, and the statistical errors
are reasonably small. The result with the mNRQCD looks encouraging, and 
so it may be worth studying the $B\to\pi l\nu$ from factors at lower $q^2$ 
using the mNRQCD formalism.

\subsubsection{$B\to D^{} l\nu$ and $B\to D^{*} l\nu$ decays}

The form factors of $B\to D^{} l\nu$ and $B\to D^{*} l\nu$ decays
can be calculated more accurately than those of
heavy-to-light decays, due to the approximate symmetry
between the initial and final states.
The branching fraction of the $B\to D^{} l \nu$ decay is given by
\bea\label{eq:decay_rateB2D}
{\rm{Br}} ( B\to D^{} l\nu  ) &=&
  |{V_{cb}}|^2  \ \int dw~F(w)^2 \times (\mbox{known factor}) ,
\eea
where $w=v_B\cdot v_D$ with
$v_{B}=p_{B}/m_B$ and $v_{D}=p_{D}/m_D$.
The form factor of the $B\to D^{} l\nu$ decay at zero recoil limit,
${{F}}_{B\rightarrow D^{}}(w=1)$,
can be precisely determined by considering the 
double ratio~\cite{Hashimoto:1999yp,Kronfeld:2000ck};
\bea
  \frac{C^{DV_0B}(t) C^{BV_0D}(t)}{C^{DV_0D}(t) C^{BV_0B}(t)}
  ~\stackrel{t\to\infty}{\rightarrow}~
  \frac{
    \langle D|{{V}}_0|B\rangle
    \langle B|{{V}}_0|D\rangle }{
    \langle D|{{V}}_0|D\rangle
    \langle B|{{V}}_0|B\rangle },
\eea
where $C^{DV_0B}(t)$ and $\langle D|{{V}}_0|B\rangle$ are respectively 
the $B\to D$ three-point function and amplitude.

\begin{figure}[t]
\begin{center}
\leavevmode
\centerline{\epsfxsize=12.5cm \epsfbox{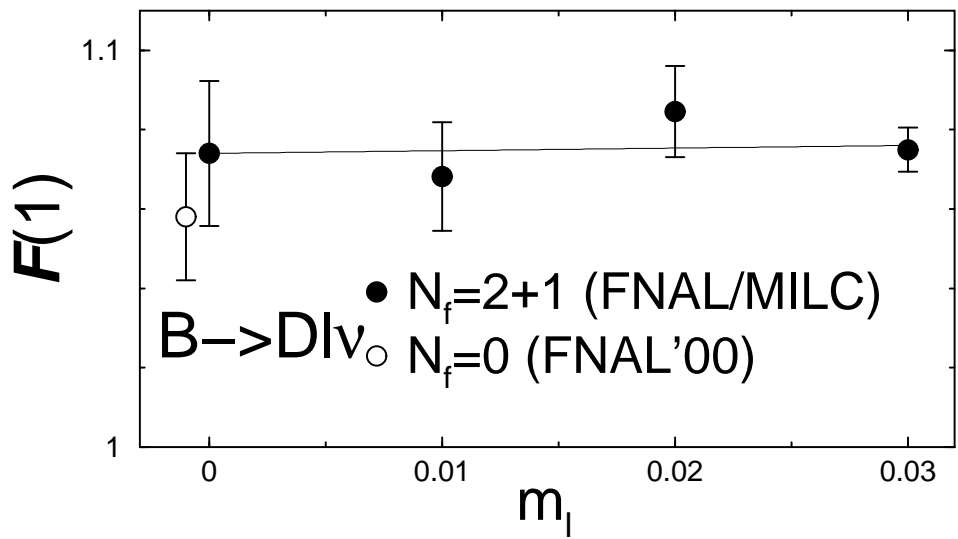}}
\caption{Light quark mass dependence of the unquenched $B\to D l\nu$ form factor at zero 
recoil~\cite{Okamoto:2004xg}.
The quenched result~\cite{Hashimoto:1999yp} is also shown around at $m_l=0$}
\label{fig:b2d}
\end{center}
\end{figure}

The Fermilab/MILC collaboration has a preliminary $n_f=2+1$ unquenched result
for the $B\to D l\nu$ form factor using the MILC configurations~\cite{Okamoto:2004xg}.
The light quark mass dependence of the form factor is mild 
and a linear chiral extrapolation was made, as shown in Fig.~\ref{fig:b2d}.
The unquenched result is ${{F}}^{B\rightarrow D}(1)=1.074(18)(16)$,
which is consistent with the quenched result~\cite{Hashimoto:1999yp}.
Combined with an average of experimental results for
$|V_{cb}|{{F}}(1)$~\cite{HFAG}, we obtain
\bea
   |V_{cb}|_\LQCDnow = 3.91(09)(34) \!\times\! 10^{-2},
\label{eq:VcbWA}
\eea
where the error from the lattice calculation (first error)
is much smaller than the experimental one (second).

For a more precise determination of $|V_{cb}|$,
the $B\to D^{*} l\nu$ decay should be used because the experimental uncertainty
is smaller. 
As for the lattice calculation, the chiral extrapolation is crucial 
because 
a singularity should appear for the form factors 
at $m_{PS}=m_{D^*}-m_{D}$ which is close to 
the physical pion mass point.
Ref.~\cite{laiho} calculated the $B\to D^{*} l\nu$ form factors in the 
staggered chiral perturbation theory (S$\chi$PT), and found that 
the singularity (seen in the continuum $\chi$PT) disappears due to 
lattice discretization effects.
The unquenched calculation of the $B\to D^{*} l\nu$ form factors is underway~\cite{laiho}.

\subsection{$K$ meson decays}

The semileptonic $K\to\pi l\nu$ decay is traditionally used to
determine $|V_{us}|$.
The form factor of $K\to\pi l\nu$ at $q^2=0$ can be precisely calculated
using the double ratio method as in the $B\to D l\nu$ case.
The first precise
lattice calculation of the $K\to\pi l\nu$ form factor
has been done in the quenched approximation
using improved Wilson quarks~\cite{Becirevic:2004ya}.
The quenched result is
\bea
f_{+}^{K\to\pi}(0) = 0.960(5)(7)~~~~~(n_f= 0),
\label{eq:f0K2pi}
\eea
which agrees with an earlier estimate using a quark model~\cite{Leutwyler:1984je}.
The uncertainty is smaller than 1\% due to the approximate 
symmetry between initial and final states,
as for the $B\to D l\nu$ form factor.
Since then, three unquenched calculations have been started.
One is the $n_f=2$ calculation using improved Wilson quarks
by the JLQCD collaboration~\cite{tsutsui}, obtaining the preliminary value
$f_{+}^{K\to\pi}(0) = 0.952(6)$.
Another $n_f=2$ calculation is underway using domain-wall quarks
by the RBC collaboration~\cite{kaneko}; their preliminary result
is $f_{+}^{K\to\pi}(0) = 0.955(12)$.
A preliminary $n_f=2+1$ calculation on the MILC configurations
is done using a combination of improved Wilson and staggered quarks 
by the Fermilab collaboration~\cite{Okamoto:2004df}, getting
$f_{+}^{K\to\pi}(0) = 0.962(6)(9)$.
These calculations rely on chiral perturbation theory to guide
the chiral extrapolation ($m_l\to m_{ud}$)
using data at $m_l \ge m_s/2$. It may be interesting to perform a direct
simulation at smaller $m_l$ using staggered fermions or chirally-improved fermions.
For technical details of the calculations, see Ref.~\cite{dawson}.
Taking a simple average of three preliminary unquenched results, I obtain
\bea
f_{+}^{K\to\pi}(0) = 0.956(12)~~~~~(n_f\ge 2).
\label{eq:f0K2pi_nf2}
\eea
Combining this with an experimental result for
$|V_{us}|f_{+}^{K\to\pi}(0)$~\cite{Alexopoulos:2004sw} gives
\bea
{|V_{us}|_{\rm semi-lep}} = 0.2264~(28)(12),
\label{eq:VusSL}
\eea
where the first error is from lattice calculations and 
the second from experiment.

$|V_{us}|$ may be determined from the leptonic
decay $K\to l\nu$~\cite{Marciano:2004uf}.
The leptonic decay constant $f_K$ has been precisely calculated in $n_f=2+1$ lattice 
QCD using improved staggered  quarks by the MILC collaboration~\cite{milc}.
Their updated result is~\cite{bernard}
\bea
f_K/f_\pi=1.198(3)(\Aerr{16}{05}).
\label{eq:fKMILC}
\eea
Using this and the experimental result for ${\rm Br}(K\to l\nu)/{\rm Br}(\pi\to l\nu)$~\cite{pdg},
one obtains
\bea
{|V_{us}|_{\rm lep}} = 0.2242(\Aerr{11}{31}),
\label{eq:VusLept}
\eea
which is consistent with that from the semileptonic decay Eq.~(\ref{eq:VusSL}).

For the Lattice 2005 value, I take a weighted average of 
Eqs.~(\ref{eq:VusSL}) and (\ref{eq:VusLept}), obtaining
\bea
|V_{us}|_\LQCDnow\!=\! \Vus\!\!\!\!\!.
\label{eq:VusWA}
\eea
There is a $2 \sigma$ disagreement between Eq.~(\ref{eq:VusWA})
and the PDG average $|V_{us}|=0.2200(26)$~\cite{pdg}.
The reason is as follows.
The PDG used the $K\to\pi l\nu$ decay for $|V_{us}|$;
the form factor is taken from Ref.~\cite{Leutwyler:1984je}
which is consistent with Eq.~(\ref{eq:f0K2pi_nf2}), but 
the $K\to\pi l\nu$ branching fraction is from earlier experimental measurements
which disagree by $\approx 2\sigma$ with the recent one used for Eq.~(\ref{eq:VusSL}).

\subsection{Other CKM magnitudes from CKM unitarity}

Having determined the 5 CKM matrix elements,
one can check a unitarity condition of the CKM matrix
using results from lattice QCD alone. From
Eqs.~(\ref{eq:VcdWA}), (\ref{eq:VcsWA}) and (\ref{eq:VcbWA}), one gets
\bea
({|V_{cd}|^2+|V_{cs}|^2+|V_{cb}|^2})^{1/2}=
{ 1.00 (10)},
\eea
which is consistent with CKM unitarity.

Conversely one may use CKM unitarity to determine
other CKM matrix elements as follows;
\bea
|V_{ud}|_\LQCDnow &=&  (1 - |V_{us}|^2 - |V_{ub}|^2)^{1/2} \nn\\
&=& \Vud, \label{eq:VudWA}\\
|V_{tb}|_\LQCDnow &=&  (1 - |V_{ub}|^2 - |V_{cb}|^2)^{1/2} \nn\\
                   &=& \Vtb, \label{eq:VtbWA}\\
|V_{ts}|_\LQCDnow &=& |V_{us}^{*}V_{ub} + V_{cs}^{*}V_{cb}|\ /\ |V_{tb}|
~\simeq~ |V_{cs}^{*}V_{cb}|\ /\ |V_{tb}| \nn\\
&=& \Vts.\label{eq:VtsWA}
\eea
One can also determine some of the Wolfenstein parameters
from Eqs.~(\ref{eq:VusWA}), (\ref{eq:VcbWA}) and (\ref{eq:VubWA}).
One gets
\bea
\lambda_\LQCDnow &=& |V_{us}| ~~~~~~~~~~=\!\!\!\! 
\Wlambda\!\!\!\!\!\!, \label{eq:Wlambda}\\
A_\LQCDnow &=& |V_{cb}|/\lambda^2 ~~~~~~=~ \WA, \label{eq:WA}\\
(\rho^2 + \eta^2)^{1/2}_\LQCDnow &=& |V_{ub}| / (A\lambda^3) ~=~ \WR.
\label{eq:WR}
\eea
The determination of remaining CKM parameters,
$(\rho,\eta)$ and $|V_{td}|$, will be discussed in the next section.

\section{CKM phase from lattice QCD}\label{sec:phase}

In this section, I extract CKM phase parameters $(\rho,\eta)$ 
and the magnitude $|V_{td}|$ using recent lattice results.
One constraint on $(\rho,\eta)$ has already been obtained
in the previous section, Eq.~(\ref{eq:WR}).
Two more constraints may be obtained from the mixing of neutral 
$B_{}$ and $K$ mesons.

The neutral $B_{}$ mixing is characterized by 
the mass difference of $B^0_{}$ and $\bar{B}^0_{}$ mesons,
which is given by
\bea
\Delta M_{B_{q}} ~=~ \mbox{(known factor)} \times f_{B_{q}}^2 B_{B_{q}}  \
{| V_{tb}^*  V_{tq}|^2 } ~~~~~(q=d,s)
\label{eq:dMB}
\eea
The nonperturbative QCD effects are contained in 
$f_{B_{q}}^2 B_{B_{q}}$, where 
$f_{B_{q}}$ is the $B_{q}$ meson decay constant defined in
an analogous way to Eq.~(\ref{eq:fDdef}) and 
$B_{B_{q}}$ is the $B_{q}$ meson bag parameter defined through
\bea
\<\bar{B}^0|(\bar{b}q)_{V-A}(\bar{b}q)_{V-A}|B^0\>
~\equiv~ \frac{8}{3}m_{B_q}\
{B_{B_{q}} f_{B_{q}}^2} ~~~~(q=d,s),
\eea
where $(\bar{b}q)_{V-A}$ is the $V-A$ current involving $b$ and $q$ quarks.
By combining the lattice calculation of 
$f_{B_{d}}^2 B_{B_{d}}$ with the precisely measured value of 
$\Delta M_{B_{d}}$, one can extract
$| V_{tb}^*  V_{td}| \simeq | V_{td}| = A \lambda^3 \sqrt{(1-\rho)^2+\eta^2}$,
which gives a constraint on $(\rho,\eta)$.
Since some uncertainties for lattice results cancel in 
the ratio $\frac{f_{B_{d}}^2 B_{B_{d}}}{f_{B_{s}}^2 B_{B_{s}}}$,
it is desirable to consider the ratio of $B_{d}$ and $B_{s}$ meson
mass differences,
\bea
\frac{\Delta M_{B_{d}}}{\Delta M_{B_{s}}}
~=~ \frac{M_{B_{d}}}{M_{B_{s}}} \ \frac{f_{B_{d}}^2 B_{B_{d}}}{f_{B_{s}}^2 B_{B_{s}}}
\ \frac{| V_{tb}^*  V_{td}|^2}{| V_{tb}^*  V_{ts}|^2} 
~\propto~ \frac{|V_{td}|^2}{| V_{ts}|^2} ~=~ 
\lambda^2 [(1-\rho)^2+\eta^2].
\eea
Up to now, however, only a lower limit is known for
$\Delta M_{B_{s}}$.
The CDF and D0 experiments are expected to measure 
$\Delta M_{B_{s}}$;
once it is available, it will provide a better constraint on $(\rho,\eta)$.

The neutral $K$ mixing is characterized by the CP-violating parameter
$\epsilon_K$, given by
\bea
|\epsilon_K| ~=~ B_K \ {\eta} [{(1- \rho)} c_1 ~+~ c_2 ]
\label{eq:eK}
\eea
where $c_1$ and $c_2$ are numerical constants, and 
$B_K$ is defined through
\bea
\<\bar{K}^0|(\bar{s}d)_{V-A}(\bar{s}d)_{V-A}|K^0\>
~=~ \frac{8}{3}m_{K}\ 
{B_{K}} f_{K}^2
\eea
with $f_{K}$ being the $K$ meson decay constant.
Combining the lattice result for $B_{K}$ and the experimental result
for $|\epsilon_K|$ gives another constraint on $(\rho,\eta)$.

As for other constrains on $(\rho,\eta)$,
recent experiments by $B$ factories
enable the measurement of all 3 angles of the CKM unitarity triangle 
$(\alpha,\beta,\gamma) = 
({\rm arg}\left( \frac{V^*_{tb}V_{td}}{V^*_{ub}V_{ud}}\right),
 {\rm arg}\left( \frac{V^*_{cb}V_{cd}}{V^*_{tb}V_{td}}\right),
 {\rm arg}\left( \frac{V^*_{ub}V_{ud}}{V^*_{cb}V_{cd}}\right))$.
In particular, the accuracy of $\beta$ is impressive.
See, for example, Refs.~\cite{abe,forti,ligeti} for recent experimental measurements of the
CKM angles.

The constraints, Eqs.~(\ref{eq:WR}), (\ref{eq:dMB}) and 
(\ref{eq:eK}), as well as other constraints, over-determine $(\rho,\eta)$.
If good precision can be achieved for each sector, 
one can test the Standard Model by seeing whether or not 
there is inconsistency between them.
Below I review recent lattice results for 
$f_{B_{}}$, $B_{B_{}}$ and $B_K$, and then
extract $(\rho,\eta)$ using them.

\subsection{$B$ meson mixing ($f_{B}^2 B_{B_{}}$)}

\subsection{$f_{B_{}}$}

Let us recall that $f_{B_{}}$ is similar to
$f_{D}$ and lattice results for the latter agree with the experimental result by CLEO-c
within $\approx$10\% uncertainties,
as seen in previous section.
This may indicate reliability of lattice results for $f_{B_{}}$.

This year the HPQCD collaboration finalized
their $n_f=2+1$ calculation of $f_{B_{}}$ using the improved staggered
quarks and the NRQCD heavy quark on the MILC 
configurations~\cite{Gray:2005ad,shigemitsu,allison,Wingate:2003gm}.
They performed simulations at the light quark masses in the range
$m_s/8 \le m_l \le m_s/2$.
The chiral extrapolations are made using various fit forms including
ones with the staggered $\chi$PT, the continuum $\chi$PT and a simple linear ansatz.
They reported that the various fits agree with each other within 3\% after the extrapolation,
suggesting that the results at the physical light quark mass 
($m_l=m_{ud}$) are insensitive
to details of the chiral fit.

\begin{figure}[t]
\vspace{1cm}
\leavevmode
    \epsfxsize=15cm
\centerline{\epsfxsize=14.cm \epsfbox{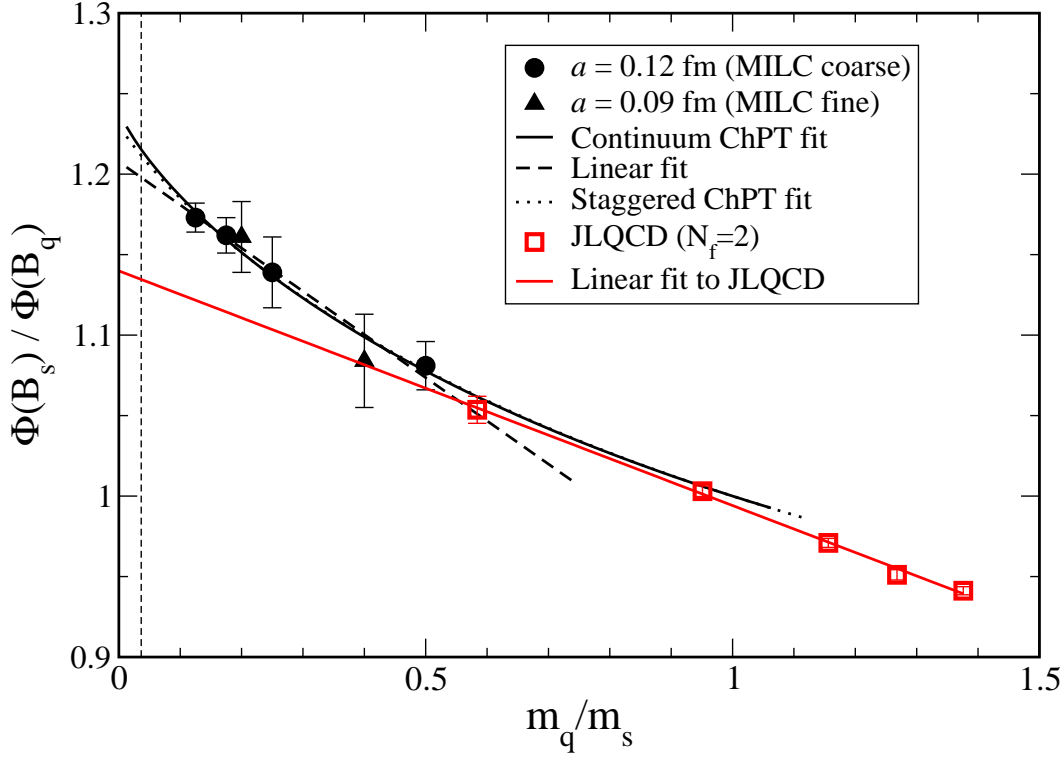}}
\caption{The light quark mass dependence of unquenched 
$f_{B_s}\sqrt{m_{B_s}}/(f_{B_d}\sqrt{m_{B_d}})$
by the HPQCD collaboration ($n_f=2+1$, black symbols)~\cite{Gray:2005ad} 
and by the JLQCD collaboration ($n_f=2$, red symbols)~\cite{Aoki:2003xb}.
Black (red) lines are the chiral extrapolations using the HPQCD (JLQCD) data.
The dashed vertical line indicates the physical light quark mass ($m_l=m_{ud}$).}
\label{fig:fBHPQCD}
\end{figure}

The $n_f=2+1$ result by the HPQCD collaboration
is shown in Fig.~\ref{fig:fBHPQCD} together with 
a $n_f=2$ result by the JLQCD collaboration~\cite{Aoki:2003xb}.
Although the JLQCD result is consistent with the HPQCD result at
$m_l\simeq m_s/2$, 
a linear chiral extrapolation
using only JLQCD data (with $m_l\ge m_s/2$)
clearly deviates from the HPQCD result
at $m_l=m_{ud}$.
I believe that this is evidence that 
data at $m_l\le m_s/2$ is required for high-precision lattice calculations.
The result by the HPQCD collaboration is~\cite{Gray:2005ad,Wingate:2003gm}
\bea
f_{B_d} &=& 216(09)(19)(07) ~\MeV,\\
f_{B_s} &=& 260(07)(26)(09) ~\MeV,
\eea
where the first errors are statistical ones and uncertainties from the chiral extrapolation,
the second are ones from the 1-loop perturbative matching for the current renormalization, 
and the third are other uncertainties.
The 2-loop or nonperturbative matchings will be 
required for an accuracy better than 5\%.
The uncertainties from the matching (and some others) cancel in
the ratio of $B_d$ and $B_s$ decay constants, giving
\bea
f_{B_s} / f_{B_d}  = 1.20 (3)(1).
\eea
This is a 3\% determination, thanks to the small error from the
chiral extrapolation.
Once $\Delta m_{B_s}$ is measured, this will significantly 
reduce the uncertainty for $|V_{ts}|/| V_{td}|$.

\begin{figure}[tb]
\begin{center}
\leavevmode
    \epsfxsize=10cm
\centerline{\epsfxsize=14.5cm \epsfbox{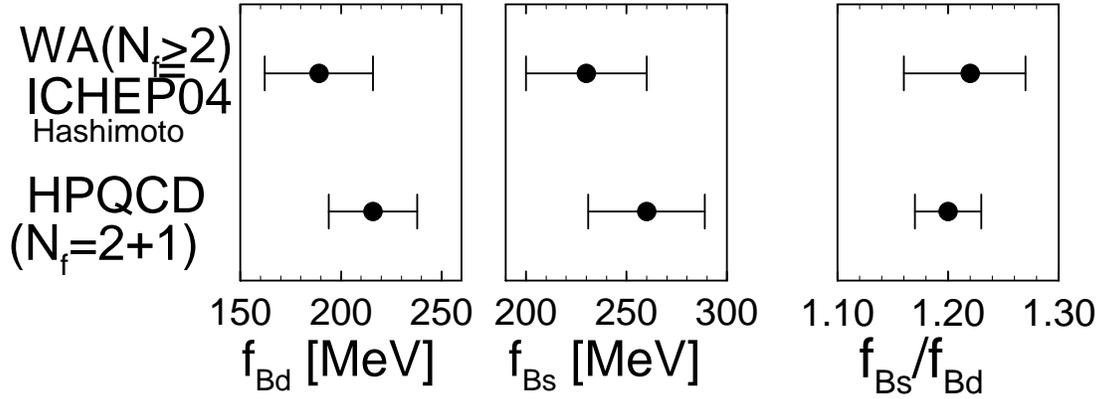}}
\caption{Comparison of the $B_{}$ meson decay constants $f_B$
in $n_f=2+1$ QCD by the HPQCD collaboration~\cite{Gray:2005ad} and 
the unquenched world average in Ref.~\cite{Hashimoto:2004hn}.}
\label{fig:compfB}
\end{center}
\end{figure}

Comparing the HPQCD results with the latest averages of unquenched results ($n_f\ge 2$)
in Ref.~\cite{Hashimoto:2004hn},
one sees a reasonable agreement 
for the individual $f_{B_d}$ and $f_{B_s}$, and 
good agreement for the ratio $f_{B_s} / f_{B_d}$, as shown in 
Fig.~\ref{fig:compfB}.
The good agreement for the ratio 
is probably due to the cancellation
of some systematic uncertainties, as mentioned above. 
I also note that the averaged value in Ref.~\cite{Hashimoto:2004hn}
is estimated by including the effect of the chiral logarithm from 
$\chi$PT to guide the chiral behavior for $m_l\le m_s/2$. 

\subsubsection{$B_{B_{}}$}

Turning to $B_{B_{}}$,
there is no new or updated result in unquenched QCD this year.\footnote{
While this paper was being completed, a new unquenched ($n_f=2$) calculation
using domain-wall light quarks and an improved static heavy quark
has been reported~\cite{Gadiyak:2005ea}. 
The results for ${B_{B_s}}/{B_{B_d}}$ and $f_{B_s}/f_{B_d} \sqrt{{B_{B_s}}/{B_{B_d}}}$
are consistent with but larger than 
Eqs~(\ref{eq:bbjlqcd}) and (\ref{eq:fBBbratio}).
}
The $n_f=2$ calculation by the JLQCD collaboration
using an improved Wilson light quark
and the NRQCD heavy quark~\cite{Aoki:2003xb}
is still only the result from unquenched QCD.
Their result is
\bea
B_{B_d}(m_b) &=& 0.836(27)(\Aerr{56}{62}) , \\
{B_{B_s}}/{B_{B_d}} &=& 1.017(16)(\Aerr{56}{17}) .
\label{eq:bbjlqcd}
\eea
Combining this and the HPQCD result for $f_{B_{}}$ gives
\bea
f_{B_d}\sqrt{\hat{B}_{B_d}} &=& 244(26) ~\MeV~,\label{eq:fbBb}\\
f_{B_s}/f_{B_d} \sqrt{{B_{B_s}}/{B_{B_d}}} &=& 1.210(\Aerr{47}{35})~,
\label{eq:fBBbratio}
\eea
where $\hat{B}_{B}$ is the renormalization group invariant
bag parameter.
These values should be compared with the previous average,
{\it e.g.,} Ref.~\cite{Hashimoto:2004hn} quoted
$f_{B_d}\sqrt{\hat{B}_{B_d}}=214(38)$ and 
$f_{B_s}/f_{B_d} \sqrt{{B_{B_s}}/{B_{B_d}}} = 1.23(6)$.
Equation~(\ref{eq:fbBb}) together with the experimental value
of  $\Delta m_{B_d}$ leads to 
\bea
|V_{td}|_\LQCDnow = \Vtd ,
\label{eq:VtdWA}
\eea
which is consistent but smaller than
the PDG value~\cite{pdg}, $|V_{td}|_{\rm PDG} = 8.3(1.6)\times 10^{-3}$.
Equation~(\ref{eq:fBBbratio}) and the forthcoming measurement of $\Delta m_{B_s}$ 
will give a 3--4\% determination of $|V_{ts}|/| V_{td}|$.

The bag parameters are also studied
in quenched ($n_f=0$) QCD using the overlap light quark action~\cite{blossier},
and in perturbation theory using the twisted mass light quark 
action~\cite{palombi,Frezzotti:2004wz}.
Using these actions has an advantage that 
operator mixings do not occur (or can be removed),
which may lead a more precise calculation of the bag parameters
in the future.
For previous quenched results, see, {\it e.g.,} 
Ref.~\cite{Yamada:2002wh}.

\subsection{$K$ meson mixing ($B_K$)}

Three new studies of $B_K$ in $n_f=2+1$ unquenched QCD are reported this 
year~\cite{gamiz,lee,cohen}.
In particular, the HPQCD collaboration presented a preliminary value using 
improved staggered quarks on the MILC configurations~\cite{gamiz},
\bea
B_K^{\bar{MS}}(2 \GeV) = 0.630(18)(130)(34),
\label{eq:bkhpqcd}
\eea
where the first error is statistical, the second is from 
the 1-loop matching which is again the largest error, and 
the third is other uncertainties.
Ref.~\cite{lee} also used an improved staggered fermion on the same configurations.
On the other hand, Ref.~\cite{cohen} used the domain wall fermion for both valence and dynamical
quarks. At present the latter two groups do not quote the physical value of $B_K$.
A preliminary study in $n_f=2$ unquenched QCD using the unimproved Wilson fermion 
is also reported in Ref.~\cite{Mescia:2005ew}.
For details of recent $B_K$ calculations, see Ref.~\cite{dawson}.

\begin{figure}[tb]
\begin{center}\label{fig:bk}
\leavevmode
    \epsfxsize=10cm
\centerline{\epsfxsize=12.5cm \epsfbox{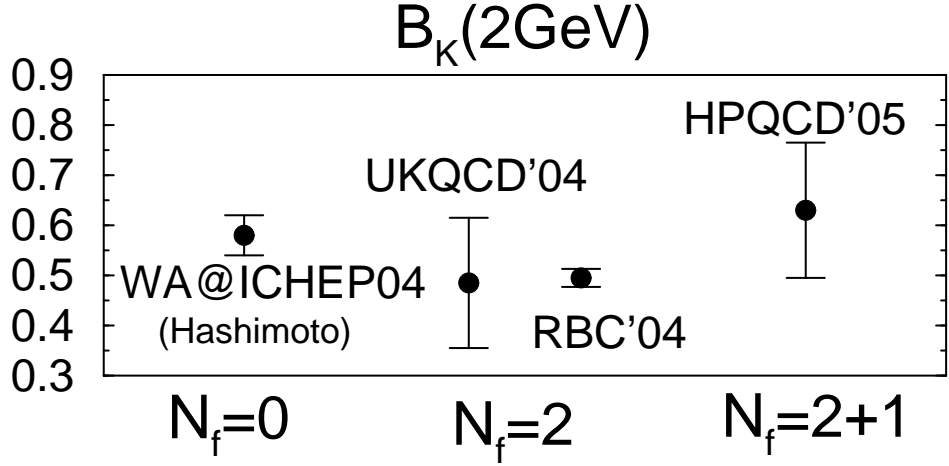}}
\caption{Comparison of recent unquenched results for $B_K$~\cite{gamiz,Flynn:2004au,Aoki:2004ht} 
and an average of 
quenched results~\cite{Hashimoto:2004hn}. 
The $n_f=2$ result by the UKQCD collaboration is obtained with an improved Wilson fermion 
action~\cite{Flynn:2004au},
and one by the RBC collaboration is obtained with a domain wall fermion action~\cite{Aoki:2004ht}.
The error for the result by the RBC collaboration is statistical only.}
\end{center}
\end{figure}

A comparison of recent unquenched results~\cite{gamiz,Flynn:2004au,Aoki:2004ht} and an average of 
quenched results~\cite{Hashimoto:2004hn} is shown in 
Fig.~13.
The unquenched results are consistent with the quenched average,
but uncertainties are much larger.
Given this, I do not take the average for the unquenched $B_K$,
and simply use Eq.~(\ref{eq:bkhpqcd}) as a representative of unquenched results
in the following unitarity triangle analysis.

\subsection{Unitarity triangle analysis}

Let us now analyze the unitarity triangle using recent lattice results to extract
$(\bar{\rho}, \bar{\eta})$.
As mentioned before, 
I use unquenched lattice results only as the theory input.
Here I adopt Eq.~(\ref{eq:WR}) for $(\rho^2 + \eta^2)^{1/2}$ from the $B\to\pi l\nu$ form factor,
Eq.~(\ref{eq:fbBb}) for $f_{B_d}\sqrt{\hat{B}_{B_d}}$, and 
Eq.~(\ref{eq:bkhpqcd}) for $B_K$.

As the experimental inputs, I use
$\Delta M_{B_d}$ and $|\epsilon_K|$.
The $B\to\pi l\nu$ branching fraction ${\rm Br}(q^2\ge 16~\GeV^2)$
is also used to obtain Eq.~(\ref{eq:WR}).
Using the lower limit of $\Delta M_{B_s}/\Delta M_{B_d}$
together with Eq.~(\ref{eq:fBBbratio}) for $f_{B_s}/f_{B_d} \sqrt{{B_{B_s}}/{B_{B_d}}}$
does not affect the result of this analysis for the reason given below.
I also use the experimental result for $\sin(2\beta)$ (=0.726(37)~\cite{HFAG}) 
from $B\to (c\bar{c}) K^{(*)}$ decays to see its impact on the
$(\bar{\rho}, \bar{\eta})$ determination.

\begin{figure}[t]
\begin{center}
\leavevmode
    \epsfxsize=10cm
\centerline{\hspace{-1cm}\epsfxsize=15.5cm \epsfbox{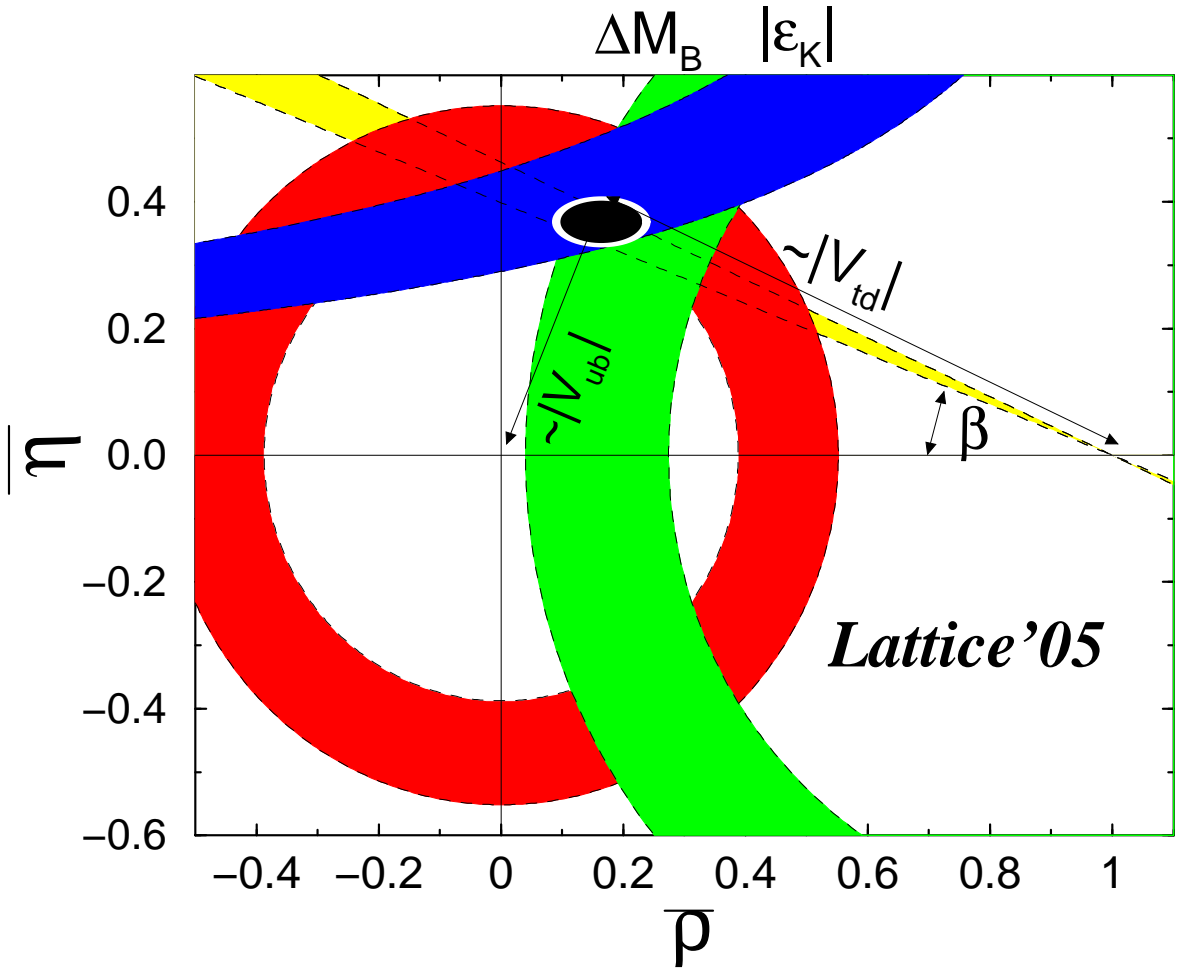}}
\vspace{-.8cm}
\caption{Unitarity triangle using recent unquenched lattice results 
for constraints from $|V_{ub}|$ (red), $\Delta M_B$ (green) and $\epsilon_K$ (blue).
The $\sin(2\beta)$ constraint (yellow) is also shown. 
The shaded regions indicate 1$\sigma$ error bands.}
\label{fig:UT}
\end{center}
\end{figure}

The unitarity triangle using lattice results together with 
the $\sin(2\beta)$ constraint is shown in Fig.~\ref{fig:UT}.
The shaded regions indicate 1$\sigma$ error bands.
A difference between this and previous ones ({\it e.g.,} one in Ref.~\cite{pdg})
is that the position of the $\Delta M_{B_d}$ bound has moved to 
the right with a smaller uncertainty.
Consequently, the bound from the lower limit of $\Delta M_{B_s}/\Delta M_{B_d}$
does not change the result for $(\bar{\rho}, \bar{\eta})$.

\begin{figure}[bt]
\begin{center}
\leavevmode
    \epsfxsize=10cm
\centerline{\epsfxsize=10.5cm \epsfbox{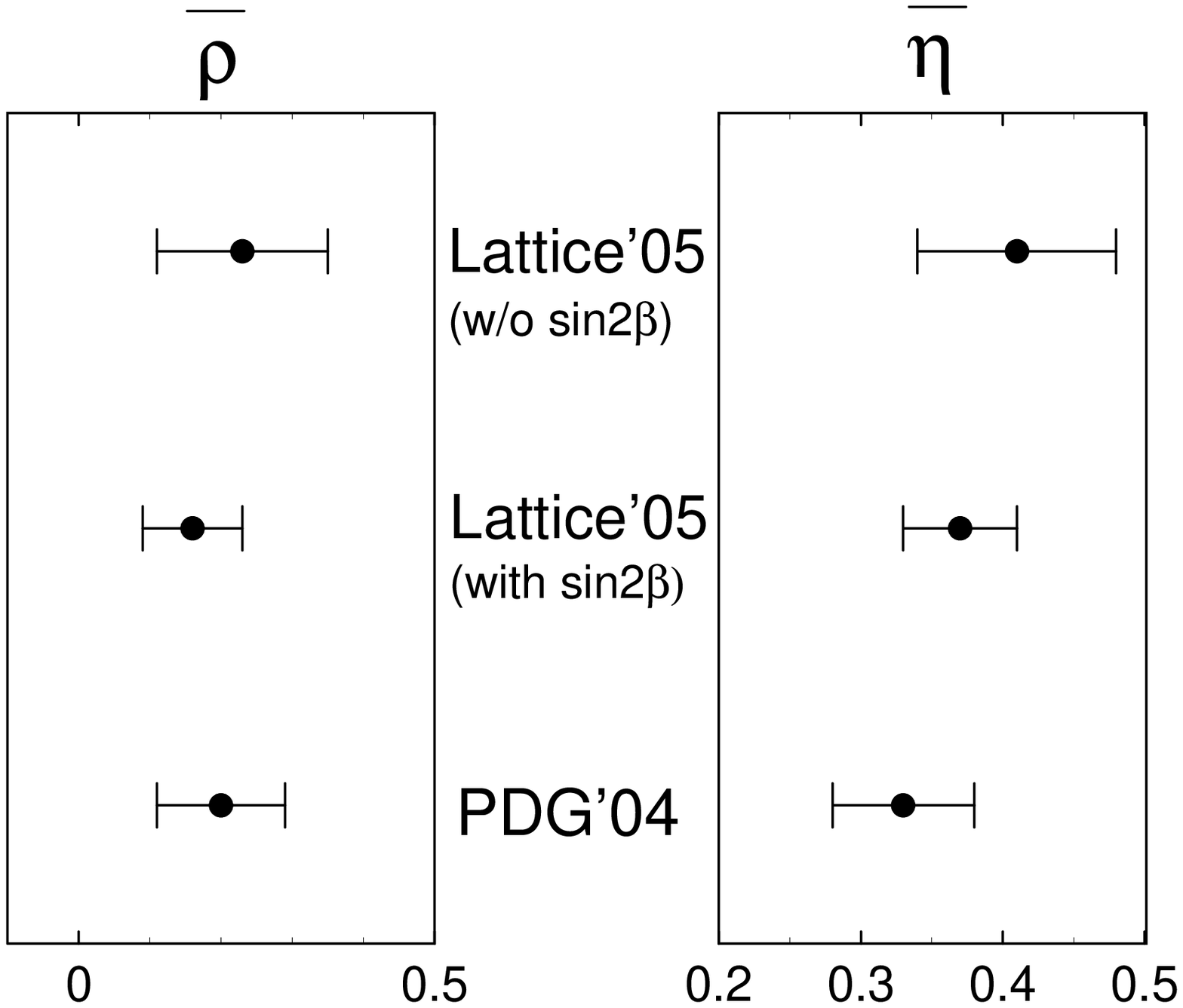}}
\caption{Comparison of $(\bar{\rho}, \bar{\eta})$ using recent unquenched 
lattice results (without and with the $\sin(2\beta)$ constraint)
and that quoted in Ref.~\cite{pdg}.}
\label{fig:rhoeta}
\end{center}
\end{figure}

From the overlapped region in $(\bar{\rho}, \bar{\eta})$ plane, I obtain
\bea
\bar{\rho} &=& 0.23(12), \\
\bar{\eta} &=& 0.41(07)
\eea 
without the $\sin(2\beta)$ constraint.
Including the  $\sin(2\beta)$ constraint significantly
improves the precision, giving
\bea
\bar{\rho}_\LQCDnow &=& \Wrho, \label{eq:rhoWA}\\
\bar{\eta}_\LQCDnow &=& \Weta, \label{eq:etaWA}
\eea 
which are consistent with the PDG values~\cite{pdg}
$\rhob=0.20(9)$ and $\etab=0.33(5)$, and the accuracy
is comparable, as shown in Fig.~\ref{fig:rhoeta}.

\section{Conclusion}\label{sec:conclusion}

This paper presents a full determination
of the CKM matrix using recent lattice results for 
gold-plated quantities.
To extract the CKM matrix elements in a uniform fashion,
results from unquenched lattice QCD are exclusively used as the theory input
for nonperturbative QCD effects.
The results for the CKM matrix elements are 
Eqs.~(\ref{eq:VcdWA}),
(\ref{eq:VcsWA}),
(\ref{eq:VubWA}),
(\ref{eq:VcbWA}),
(\ref{eq:VusWA}),
(\ref{eq:VudWA}),
(\ref{eq:VtbWA}),
(\ref{eq:VtsWA}) and 
(\ref{eq:VtdWA}).
The results for the Wolfenstein parameters are 
Eqs.~(\ref{eq:Wlambda}),
(\ref{eq:WA}),
(\ref{eq:rhoWA}) and 
(\ref{eq:etaWA}).
These are summarized in one place, Eqs.~(\ref{eq:WACKM})-(\ref{eq:WAend}).

At present, many unquenched results are obtained with 
improved staggered fermion actions.
On the other hand,
fewer unquenched results are obtained with other fermion formalisms
(such as the Wilson-type fermion, domain wall fermion and overlap fermion),
especially for the $n_f=2+1$ case.
Consequently, the results for the CKM matrix elements presented here are often
estimated from only one or two unquenched calculations.
I expect that more unquenched results using other lattice fermions will appear
in the near future, 
leaving future reviewers to make a more serious average of the CKM matrix.

The unquenched results for $D$ physics (such as the $D\to\pi l\nu$ form factor
$f_+^{D\to\pi}$ and 
the $D$ meson decay constant $f_D$) are in agreement with recent experimental results.
This may increase confidence in lattice results for similar quantities for 
$B$ physics (such as the $B\to\pi l\nu$ form factor and $f_B$).

The typical accuracy of most of gold-plated quantities is $O(10\%)$,
being dominated by uncertainties from
the lattice discretization effects, perturbative matching, and 
the chiral extrapolation.
The last uncertainty especially applies to lattice fermions other than 
the staggered fermion.
To achieve an accuracy better than 5\%,
simulations at smaller lattice spacings and smaller quark masses
and higher-order matchings will be required. 

\begin{figure}[tb]
\begin{center}
\leavevmode
    \epsfxsize=10cm
\centerline{\hspace{0cm}\epsfxsize=12cm \epsfbox{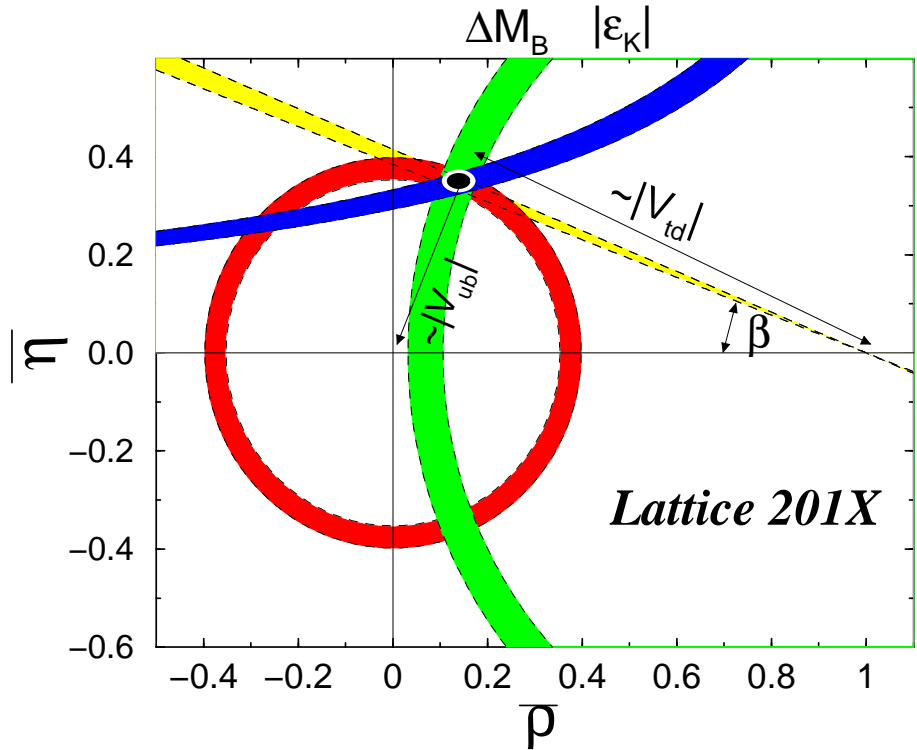}}
\vspace{-.8cm}
\caption{Expected unitarity triangle
with 5\% accuracies for the $B\to\pi l\nu$ form factor and $B_K$
and a 3\% accuracy for $f_{B_s}/f_{B_d}\sqrt{B_{B_s}/B_{B_d}}$.
The $\Delta M_{B_s}$ measurement is also assumed here.}
\label{fig:UTfuture}
\end{center}
\end{figure}

A better accuracy (3\% or less) is obtained for the $B\to D^{} l\nu$
and $K \to\pi l\nu$ form factors due to the approximate symmetry between initial and final states,
and for the decay constant ratio $f_{B_s}/f_{B_d}$ due to the cancellation of 
systematic errors. The latter will lead to a more precise constraint on
$(\rho,\eta)$ once $\Delta M_{B_s}$ is measured by experiment.

Assuming the $\Delta M_{B_s}$ measurement and 5\% (or better) accuracies for lattice results,
the unitarity triangle will be something like Fig.~\ref{fig:UTfuture}.
I hope that this will be realistic in next 5 years
so that we can more precisely test the standard model using lattice QCD
and be ready for new physics.

\section*{Acknowledgment}

I wish to thank 
S.~Aoki,
M.~Artuso,
C.~Bernard,
C.~Davies,
C.~Dawson,
Steven~Gottlieb,
E.~Gulez,
S.~Hashimoto,
T.~Kaneko,
A.~Kronfeld,
J.~Laiho,
P.~Mackenzie,
H.~Matsufuru,
M.~Nobes,
T.~Onogi,
G.~Rossi,
J.~Shigemitsu,
J.~Simone,
S.~Tamhankar,
and
N.~Yamada
for useful communications.
I am grateful to
A.~Kronfeld, J.~Shigemitsu and A.~Akeroyd 
for a careful reading and for valuable suggestions on the manuscript.
Finally I thank the Lattice 2005 organizers
for inviting me to the conference.

\end{document}